\documentclass[11pt]{article}
\usepackage{geometry}
\usepackage{graphicx}
\usepackage{amsmath}
\usepackage{nccmath}
\usepackage{amssymb}
\usepackage{caption}
\usepackage{subcaption}
\usepackage{multirow}
\usepackage{array}
\usepackage{cite} 
\usepackage{xcolor}  
\usepackage{hyperref}

\hypersetup{
	colorlinks=true,        
	linkcolor=blue,         
	citecolor=green!60!black, 
	urlcolor=magenta,     
	filecolor=red,         
	anchorcolor=cyan,      
	pdfborder={0 0 0}      
}
\usepackage{titlesec}
\titleformat{\section}{\large\bfseries}{\thesection}{1em}{}
\titleformat{\subsection}{\normalsize\bfseries}{\thesubsection}{1em}{}
\providecommand{\keywords}[1]{\small \textbf{\textit{Keywords---}} #1}
\makeatletter
\def\maketitle{
	\begin{center}
		{\LARGE \bfseries \@title}\par
		\vskip 0.5em  
		{\large \@author}\par
		\vskip 0.2em 
	\end{center}
	\thispagestyle{plain}  
	\setcounter{footnote}{0} 
}
\makeatother

\title{\large \textbf{Exact Solution of Schrödinger equation for Complex Mass Quantum System under Complex Morse Potential to study emergent matter types}}
\author{\normalsize Partha Sarathi\textsuperscript{1*‡} 
	\hspace{-0.4em} \normalsize and
	Bhaskar Singh Rawat\textsuperscript{2 $\dagger$ §} \\
	\vspace{0.5em}
	\footnotesize \textsuperscript{1}Department of Physics, Maharaja Agrasen College, University of Delhi, Vasundhara Enclave, Delhi-110096, India \\
	\footnotesize \textsuperscript{2}Department of Physics, Hemvati Nandan Bahuguna Garhwal University, Srinagar-246174, Uttarakhand, India
}
\date{} 

\begin{document}
	
	\maketitle
	\vspace{-0.5em} 
	
\renewcommand{\thefootnote}{\fnsymbol{footnote}}

	\footnotetext[1]{\href{mailto:parthasarthi@mac.du.ac.in}{parthasarthi@mac.du.ac.in}}
	\footnotetext[2]{\href{mailto:bsrawat.phy@gmail.com}{bsrawat.phy@gmail.com}}
	
	\footnotetext[3]{\href{https://orcid.org/0000-0002-0061-2061}{0000-0002-0061-2061 (P. Sarathi)}} 
	\footnotetext[4]{\href{https://orcid.org/0009-0009-4527-2601}{0009-0009-4527-2601 (B. S. Rawat)}} 
	
\renewcommand{\thefootnote}{\arabic{footnote}}

		\begin{abstract}
		We present exact solutions of the Schrödinger equation for a quantum system with complex mass subjected to a complex Morse potential in the extended complex phase space. The normalized eigenfunctions and corresponding eigenspectra are derived within a non-Hermitian framework, ensuring consistent probability densities. Conditions for the reality of the spectra are established and used to analyze the dependence of eigenvalue behaviour on potential parameters. The study reveals distinct regimes of spectral characteristics arising from the interplay of complex mass, the Morse parameter, and eigenvalues, leading to the emergence of five classes of non-Hermitian quantum states. By analysing the energy eigenspectra, normalization conditions, and probability density profiles across parameter space, we identify regimes corresponding to real-spectrum Hermitian-like matter, quasi-stable or resonant states, purely complex quantum matter, non- physical, non-normalizable states, and a quasi-classical determinate regime in which the probability density becomes spatially static. One of these system exhibits a non-dissipative, collisionless state with long-range gravitational-like characteristics, suggesting a theoretical analogue for dark matter within a non-Hermitian quantum framework. Further, the five identified classes of matter may be interpreted as distinct matter phases of a single quantum system governed by complex mass and Morse parameters This classification elucidates the boundary between physical and non-physical regimes in complex quantum systems and provides a unified approach for interpreting stability, resonance, and emergent classicality arising from complex parameters.

    	\end{abstract}
	
	\noindent\rule{\linewidth}{0.4pt}
	
	\section{Introduction} \label{sec1}
	
	Conventional quantum mechanics is built on the premise that observables correspond to Hermitian operators, ensuring real eigenvalues and unitary, probability-conserving dynamics. Yet, over the past few decades, non-Hermitian Hamiltonians have emerged as powerful, effective tools for describing and engineering a variety of phenomena in open and non-conservative systems. A central breakthrough came with the study of parity–time (PT) symmetry, introduced by Bender and collaborators, who showed that certain non-Hermitian Hamiltonians can nevertheless exhibit entirely real spectra and unitary evolution under suitably defined inner products \cite{bender2007making, bender1999PT, bender2005introduction, bender2004extension, bender1998real, mostafazadeh2010pseudo}. This insight has laid the groundwork for effective non-Hermitian descriptions across photonics, condensed matter, nuclear and molecular systems, and even cosmology.
	
	A particularly fertile area of development has been photonics, where the isomorphism between the paraxial diffraction equation and the Schrödinger equation enables mapping optical gain and loss to complex refractive indices. This has led to PT-symmetric realizations in coupled waveguides and resonators, giving rise to phenomena such as unidirectional invisibility, laser- absorber devices, asymmetric transport, single-mode lasing, and enhanced sensing \cite{takata2017pt, ozdemir2019parity, longhi2017parity}. Beyond optics, effective non-Hermitian Hamiltonians have been applied in magnonic systems, hybrid quantum devices, and condensed matter platforms, revealing features such as non-Hermitian topology, dynamical phase transitions, non-Hermitian skin effects, and the role of exceptional points (EPs) in steady-state and dynamical responses \cite{miri2019exceptional, ding2022non, li2024observation, lin2022topological, hurst2022non, lin2011unidirectional, feng2014single}. EPs, in particular, have been experimentally realized in microcavities \cite{lee2009observation}  and photonic crystals \cite{takata2021observing}, and are known to induce chiral dynamics and sensitivity enhancements \cite{lee2009observation, takata2021observing, chen2017exceptional, qin2026photonic}. Moreover, engineered non-Hermitian photonic structures have demonstrated coherent perfect absorption and laser emission through anisotropic emissions \cite{li2023coherent}. Important progress has also been reported in non-Hermitian lattice systems. Zhu, Lü, and Chen (2014) analyzed a PT-symmetric extension of the Su–Schrieffer–Heeger (SSH) model with complex boundary potentials, showing how PT symmetry governs the transition between unbroken and broken spectral phases \cite{zhu2014pt}. Experiments on coupled-cavity arrays have further demonstrated spontaneous PT-symmetry breaking in realistic photonic architectures with balanced gain and loss0\cite{xing2017spontaneous}. Complementing these, Jin and Song (2009) provided exact solutions of PT-symmetric tight-binding chains and mapped them to Hermitian counterparts, clarifying the equivalence between non-Hermitian and Hermitian descriptions in specific regimes \cite{jin2009solutions}. Collectively, these studies highlight how non-Hermitian PT-symmetric Hamiltonians enrich the fundamental landscape of quantum theory while paving practical pathways for new condensed matter and photonic phenomena.
	
	Given their admissibility to real eigenspectra, PT-symmetric Hamiltonians have critically influenced the literature. However, RS Kaushal and Parthasarathi (2002), by implementing an extended complex phase-space approach, exactly solved various non-Hermitian Hamiltonian systems in one dimension and studied the possible conditions under which the various complex potentials possess real spectra \cite{kaushal2002quantum}. Further, the extended complex phase-space approach had been implemented to solve for quantum problems such as the bicomplex oscillator \cite{bagchi2015bicomplex}, quantum motion \cite{yang2005quantum, yang2007quantum}, supersymmetric quantum mechanics \cite{parthasarathi2004complex}, PT and non-PT symmetry \cite{bhardwaj2016quantum}, three- dimensional harmonic potentials \cite{bhardwaj2014eigenspectra}, sextic potentials \cite{parthasarathi2003quantum, chand2009solution}, deictic, quadratic and Octic potentials \cite{singh2009solution, singh2014solving, singh2013quantum}.
	
	Parallel to these advances, exactly solvable models have provided deep insight into spectral structures of non-Hermitian systems. Analytical solutions for PT symmetric potentials—via factorization, SUSY-QM, or the Nikiforov–Uvarov method—yield closed-form eigenvalues and eigenfunctions, offering controlled platforms to explore real-to-complex spectral transitions \cite{ahmed2001real, bagchi2000sl, znojil1999pt, dutt1988supersymmetry}. Despite the successful applications of the aforementioned methods for various PT symmetric potentials, the complex Morse potential is comparatively challenging to treat analytically. The challenge arises due to the non-linear exponential form of the potential, making  traditional methods mathematically intricate. One approach is to study the system in an extended complex phase space, reducing the challenging aspect of the problem. The complex Morse potential has applications in molecular physics (vibrational spectra of diatomics), nuclear and condensed matter physics (effective interaction potentials), and optics (gain–loss photonic lattices) \cite{ahmed2001real, bagchi2000sl, znojil1999pt, dutt1988supersymmetry}. Its tractability also makes it a natural laboratory for probing spectral singularities, exceptional points, and biorthogonal eigenfunction frameworks \cite{miri2019exceptional}.
	
	An important conceptual extension within non-Hermitian quantum mechanics is the introduction of complex masses, which naturally arise in systems with dissipation, decay, or amplification. In high-energy physics, complex masses appear in the description of unstable resonances, where the imaginary component corresponds to the decay width and leads to the familiar Breit–Wigner propagators \cite{weisskopf1930berechnung}. The unstable states such as mesons and hadronic resonances are often described with a complex mass, where the imaginary component is proportional to the decay width. In condensed matter and photonics, effective quasiparticles such as magnons, polaritons, or optical modes can acquire complex masses through environmental coupling, providing a compact description of loss, gain, and finite lifetimes in non-Hermitian band structures \cite{miri2019exceptional, hurst2022non}. Complex masses have also been employed in nuclear and molecular physics, where models such as the complex Morse potential incorporate non-Hermiticity to capture bound states with decay or resonance features \cite{lin2011unidirectional, feng2014single, li2023coherent}. More recently, they have been invoked in cosmology and dark-sector physics, including phantom cosmology models \cite{mostafazadeh2020time}, non-Hermitian quantum cosmology \cite{guo2024non}, and dissipative dark-matter scenarios \cite{mannheim2006alternatives, bai2024correlators}. In cosmology, position- dependent or complex masses have been proposed for scalar fields driving dark energy or phantom cosmology. Similarly, in dark-matter model building, complex mass terms can represent semi-stable or dissipative dark-sector states, where decay into hidden channels modifies early-universe dynamics. These ideas extend to neutrino oscillations in a dark-matter background, where effective non-Hermitian Hamiltonians with complex mass-like parameters encode absorption and decoherence. Collectively, these applications show that complex masses provide a unifying language for treating unstable or non-equilibrium dynamics across fields as diverse as particle physics, condensed matter, optics, and cosmology.
	
	Complex masses provide a testing ground for non-Hermitian extensions of solvable systems, enabling the exploration of spectral reality and normalized eigenfunctions admitted by such systems. This was explicitly demonstrated in the context of the complex Morse potential, showing how both constant and position-dependent complex masses yield real spectra under suitable constraining conditions among potential parameters \cite{sarathi2021application, sarathi2025exact}. The exact solutions of the Schrödinger equation for the complex Morse potential under negative mass conditions were obtained, demonstrating the possibility of real bound states from unconventional mass frameworks \cite{sarathi2021application}. More recently, these solutions were generalized to systems with position- dependent complex mass, thereby extending the applicability of the model \cite{sarathi2025exact}. Together, these studies demonstrate how exactly solvable non-Hermitian models such as the complex Morse potential can bridge formal mathematical advances with applications in molecular physics, negative-mass systems, and even cosmological and dark-matter contexts. In this paper, we build upon these foundations to advance the role of non-Hermitian quantum mechanics as a unifying framework for both fundamental theory and emergent physical phenomena.
	
	Non-Hermitian methods have also found several applications in cosmology and high-energy physics. PT-symmetric complex Lagrangians have been invoked to model transitions from standard dark energy to phantom regimes \cite{andrianov2006complex}. Time-dependent pseudo-Hermitian Hamiltonians provide a geometric framework for quantum cosmology, where evolving Hilbert spaces reflect the dynamics of the expanding universe \cite{mostafazadeh2020time}. Holographic approaches, via the thermofield- double (TFD) formalism, link non-Hermitian operators to black hole states and early-universe correlators \cite{de2009non}. Extensions of non-Hermitian QFT respecting Poincaré invariance have been explored for cosmological model building \cite{guo2024non}, and even astrophysical phenomena such as high- energy cosmic-ray propagation have been discussed within non-Hermitian frameworks \cite{gomes2024non}.
	
	Another exciting frontier is the interplay between non-Hermitian quantum mechanics and dark matter. Non-Hermitian effective Hamiltonians provide natural descriptions of neutrino oscillations through dark matter backgrounds, incorporating absorption, decoherence, and CP- asymmetric effects \cite{ohlsson2016non}. Pseudo-Hermitian QFT offers consistent frameworks for dark sectors with decay widths or dissipative interactions \cite{chernodub2025pseudoreal}. At the gravitational level, Mannheim has argued that PT-symmetric extensions of quantum mechanics could account for galactic rotation curves and lensing without invoking particle dark matter \cite{mannheim2006alternatives}. Extensions to cosmological perturbation theory allow for non-Hermitian treatments of primordial correlators \cite{bai2024correlators}, while NH- inspired gauge portals suggest new mechanisms for thermal freeze-out and collider signatures \cite{rizzo2025toward}. Finally, the extreme sensitivity at exceptional points has been highlighted as a potential route to enhanced laboratory detection of light dark matter and dark photons \cite{mcdonald2020exponentially}.
	
	In this paper, we aim to find the exact solution of the Schrodinger equation for the Complex Morse potential and investigate distinct regimes of spectral characteristics arising from the interplay of complex mass, the Morse parameter, and eigenvalues, leading to the emergence of several classes of non-Hermitian quantum states. The Schrodinger equation is exactly solved for particles characterized by complex mass under action of Complex Morse potential
	
	$$V(x) = V_0\left[e^{-2ax}-2e^{-ax}\right].$$
	
	where $V_0$ is the well depth, $x$ is the internuclear distance (bond length), $x>0$ and $a$ is related to the vibrational constant $\mu$ as
	
	$$a = \pi \mu \sqrt{\frac{2 M}{V_0}}.$$
	
	The reduced mass $M$ of the system is considered a complex mass and is a function of the internuclear distance $x$ which is taken as complex. In general, the parameters $V_0$ and $a$ are also considered complex as the objective of the study is to investigate the bonding between two particles having complex mass.
	
	The arrangement of the paper is as follows: In Section \ref{sec2}, a general formulation for the solution of the Schrödinger equation for a general class of complex potential is enumerated. In Section \ref{sec3}, the exact solution of the Schrödinger equation is obtained for the general class of one-dimensional complex Morse potential with complex mass and its eigenvalues and eigenfunction are computed. The normalized eigenfunction and probability density is derived in Section \ref{sec4}. The eigenvalues and normalized eigenfunction are analyzed in Section \ref{sec5} by plotting the same with respect to the imaginary part of the mass function and Morse parameter. The admissibility of real eigenvalues for the Complex Morse potential is discussed in Section \ref{sec6} and the nature of the solution of the Reality of Spectrum (ROS) case is investigated. Lastly, a general discussion on the results and applications of such studies is presented in Section \ref{sec7} with efforts to enumerate and classify several forms of possible matter arising from Complex Morse potential based on mass and potential parameters.
	
	\section{Schrödinger Equation in the Extended Complex Phase Space} \label{sec2}
	
	To investigate the characteristics of a diatomic molecule defined under the influence of a complex Morse potential and a complex mass, we adopt the extended complex phase space (ECPS) formalism. This approach provides a significant analytical advantage, as it allows the application of complex analysis tools, particularly the Cauchy–Riemann (CR) conditions to explore the behaviour of non-Hermitian quantum systems with complex potentials.
	
	In this framework, the conventional phase space variables ($x$, $p$) are reformulated into complex canonical pairs as follows:
	
	\begin{equation}
		x = x_1 + ip_2, \qquad p = p_1 + ix_2. \label{eq1}
	\end{equation}
	
	where ($x_1$, $p_1$) and ($x_2$, $p_2$) are canonical conjugate pairs. This transformation effectively doubles the phase space, enabling a simultaneous treatment of the real and imaginary components of the physical variables.
	
	The corresponding non-Hermitian Schrödinger equation in ECPS takes the form:
	
	\begin{equation}
		\hat{H}(x,p)\psi(x) = E\psi(x).	\label{eq2}		
	\end{equation}
	
	where $H(x, p)$ denotes the non-Hermitian Hamiltonian operator, expressed as:
	
	\begin{equation}
		H(x,p) = -\frac{\hbar^2}{2m}\frac{\partial^2}{\partial{x^2}} + V(x). \label{eq3}
	\end{equation}
	
	In this representation, the quantum momentum operator is generalized as:
	
	$$p \rightarrow -i\hbar \frac{\partial}{\partial x}.$$
	
	which implies the transformations
	
	$$p_{1} \rightarrow -\frac{\hbar}{2} \frac{\partial}{\partial p_{2}}, \quad \text{and} \quad x_{2} \rightarrow -\frac{\hbar}{2} \frac{\partial}{\partial x_{1}}.$$
	
	The complex eigenfunction  is assumed to be a function of the extended variables ($x_1$, $p_2$) and can be expressed as:
	
	\begin{equation}
		\psi(x) = \psi_r(x_1,p_2) +i\psi_i(x_1,p_2). \label{eq4}
	\end{equation}
	
	where $\psi_r$ and $\psi_i$ denote the real and imaginary components of the wavefunction, respectively.
	
	Assuming complex-valued system parameters:
	
	\begin{equation}
         m = m_{r} + im_{i}, \quad V(x) = V_r(x_1,p_2) +iV_i(x_1,p_2), \quad E = E_r + iE_i. \label{eq5}
	\end{equation}
	
	and applying the Cauchy–Riemann conditions to maintain analyticity in the ECPS, we obtain:
	
	\begin{equation}
		\frac{d \psi_r}{d x_1} = \frac{d \psi_i}{d p_2}; \quad \frac{d \psi_r}{d p_2} = -\frac{d \psi_i}{d x_1}. \label{eq6}
	\end{equation}
	
	Substituting equations (\ref{eq4})-(\ref{eq6}) into the Schrödinger equation (\ref{eq2}) yields a pair of coupled partial differential equations (PDEs) governing $\psi_r$ and $\psi_i$
	
	\begin{subequations} \label{eq7}
		\begin{align}
			\frac{\hbar^2}{2|m|^2}\left[m_r \psi_r'' + m_i \psi_i''\right] &= \left(V_r - E_r\right) \psi_r - \left(V_i -E_i\right) \psi_i, \label{eq7a}\\
			\frac{\hbar^2}{2|m|^2}\left[m_r \psi_i'' - m_i \psi_r''\right] &= \left(V_i - E_i\right) \psi_r + \left(V_r -E_r\right) \psi_i. \label{eq7b}
		\end{align}
	\end{subequations}
	
	where $\psi'' = \frac{d^2 \psi}{d x_{1}^2}$ and $|m|^2 = m_{r}^2 + m_{i}^2$  denotes the modulus  of the complex mass.
	
	To solve the above equations, we employ the eigenfunction-ansatz method, assuming a ground state solution of the form:
	
	\begin{equation}
		\psi(x) = e^{i g(x)} \label{eq8},
	\end{equation}
	
	where $g(x)$ is a complex function given by $g(x) = g_{r}(x_1, p_2) + i g_{i}(x_1, p_2)$. Substituting equation (\ref{eq8}) into (\ref{eq4}), we obtain the real and imaginary components:
	
	\begin{equation}
		\psi_r = e^{-i g_i} \cos{g_r}; \quad \psi_i = e^{-g_i}  \sin{g_r} \label{eq9},
	\end{equation}
	
	Replacing these expressions into equations (\ref{eq7a}) and (\ref{eq7b}) yields a set of coupled differential equations in terms of $g_r$ and $g_i$:
	
	\begin{subequations} \label{eq10}
		\begin{align}
			\frac{\hbar^2}{2|m|^2}\left[m_r \left(-g_r'' + 2g_r' g_i' \right) + m_i \left(g_i'^2 - g_r'^2 -g_i'' \right) \right] &= E_i -V_i, \label{eq10a}\\
			\frac{\hbar^2}{2|m|^2}\left[m_r \left(g_i'^2 - g_r'^2 -g_i'' \right) - m_i \left(-g_r'' + 2g_r' g_i' \right) \right] &= V_r - E_r. \label{eq10b}
		\end{align}
	\end{subequations}
	
	Here, primes denote differentiation with respect to $x_1$.
	
	Equations (\ref{eq10a}) and (\ref{eq10b}) form the underlying analytic framework to determine the eigenvalues ($E_r$, $E_i$) and the eigenfunction of the system under the combined influence of complex mass and complex Morse potential.
	
	\section{Eigenvalue and Eigenfunction } \label{sec3}
	
	In this section, we focused on solving Schrödinger Equation, given in Eq. (\ref{eq2}), for the complex Morse potential 
	
	\begin{equation}
		V(x) = V_o \left[ e^{-2ax} - 2 e^{-ax}\right] \label{eq11}
	\end{equation}
	
	where $V_o$, $a$ are also complex, i.e $V_o = V_{or} + i V_{oi}$, $a = a_r + i a_i$. In an extended complex phase space, the imaginary and real parts of the potential can be given as, 
	
	\begin{subequations} \label{eq12}
		\begin{align}
			V_r(x_1,p_2)  &= V_{or}\left[e^{-2X}\cos{2Y}-2e^{-X}\cos{Y}\right]+V_{oi}\left[e^{-2X}\sin{2Y}-2e^{-X}\sin{Y}\right], \label{eq12a} \\
			V_i(x_1,p_2)  &= V_{oi}\left[e^{-2X}\cos{2Y}-2e^{-X}\cos{Y}\right]-V_{or}\left[e^{-2X}\sin{2Y}-2e^{-X}\sin{Y}\right]. \label{eq12b}
		\end{align}
	\end{subequations}
	
	The values of $X$ and $Y$ are  $X = a_r x_1 - a_i p_2; \quad Y = a_i x_1 + a_r p_2$. Further, the ansatz of the eigenfunction is considered as,
	
	\begin{subequations} \label{eq13}
		\begin{align}
			g_r(x_1,p_2) &= \beta_1x_1-\alpha_1p_2+\beta_3e^{-X}\cos{Y}, \label{eq13a} \\
			g_i(x_1,p_2) &= \alpha_1x_1+\beta_1p_2-\beta_3e^{-X}\sin{Y}. \label{eq13b}
		\end{align}
	\end{subequations}
	
	Substituting the ansatz defined in Eq. (\ref{eq13a}) and (\ref{eq13b})  and potential real and imaginary parts, Eq. (\ref{eq12a}) and (\ref{eq12b}) in the derived coupled equations (\ref{eq10a}) and (\ref{eq10b}), one can get the following set of non-repeating  equations by comparing coefficients of $e^{-X} \cos{Y}, \quad e^{-X} \sin{Y}, \quad e^{-2X} \cos{2Y} \quad \text{and} \quad e^{-2X} \sin{2Y}$.
	
	\begin{subequations} \label{eq14}
		\begin{align}
			E_r &= \frac{\hbar^2}{2|m|^2}\left[m_r (\beta_1^2 - \alpha_1^2) + m_i (2 \alpha_1 \beta_1)\right], \label{eq14a} \\
			E_i &= \frac{\hbar^2}{2|m|^2}\left[m_r (2 \alpha_1 \beta_1) - m_i (\beta_1^2 - \alpha_1^2)\right], \label{eq14b} \\
			V_{or} &= \frac{- \hbar^2 \beta_3}{4|m|^2} \left[m_r \left\{2 \beta_1 a_r - 2 \alpha_1 a_i - 2 a_r a_i\right\} + m_i \left\{(a_r^2-a_i^2)+ 2\alpha_1 a_r + 2 \beta_1 a_i\right\}\right], \label{eq14c} \\
			V_{oi} &= \frac{- \hbar^2 \beta_3}{4|m|^2} \left[m_r \left\{(a_r^2-a_i^2)+ 2\alpha_1 a_r + 2 \beta_1 a_i\right\} - m_i \left\{2 \beta_1 a_r - 2 \alpha_1 a_i - 2 a_r a_i\right\}\right], \label{eq14d} \\
			V_{oi} &= \frac{\hbar^2 \beta_3^2}{2 |m|^2}\left[m_r \left\{2 a_r a_i\right\}-m_i \left\{a_r^2-a_i^2\right\}\right], \label{eq14e} \\
			V_{or} &= \frac{\hbar^2 \beta_3^2}{2 |m|^2}\left[m_r \left\{a_r^2-a_i^2\right\}+ m_i \left\{2 a_r a_i\right\} \right], \label{eq14f} 
		\end{align}
	\end{subequations}
	
	The value of $\beta_3$ can be obtained from Eq. (\ref{eq14e}) and (\ref{eq14f}) as
	
	\begin{equation}
		\beta_3 = \left[\frac{2 |m|^2 V_{or}}{\hbar^2 \left[m_r (a_r^2-a_i^2) + 2 m_i a_r a_i\right]}\right]^{\frac{1}{2}}. \label{eq15}
	\end{equation}
	
	The constraining relation between the real and imaginary part of the dissociation energy from Eq. (\ref{eq14e}) and (\ref{eq14f}) can be given as,
	
	\begin{equation}
		\frac{V_{oi}}{V_{or}} = \frac{2 m_r a_r a_i - m_i (a_r^2 - a_i^2)}{m_r (a_r^2 - a_i^2)+ 2 m_i a_r a_i}. \label{eq16}
	\end{equation}
	
	Further, the value of $\alpha_1$ and $\beta_1$ in terms of $\beta_3$ can be expressed by solving Eq. (\ref{eq14c}) and (\ref{eq14c})
	
	\begin{subequations} \label{eq17}
		\begin{align}
			\alpha_1 &= \beta_3 a_i - \frac{a_r}{2}, \label{eq17a} \\
			\beta_1 &= \beta_3 a_r + \frac{a_i}{2}. \label{eq17b}
		\end{align}
	\end{subequations}
	
	Using Eq. (\ref{eq17a}) and (\ref{eq17b}), the expression of the real and imaginary parts of the eigenvalue is given as, 
	
	\begin{subequations} \label{eq18}
		\begin{align}
			E_r &= \frac{\hbar^2}{2|m|^2} \left[m_r\left\{(a_r^2-a_i^2)(\beta_3^2 - \frac{1}{4}) + 2\beta_3 a_r a_i\right\} + m_i \left\{2 a_r a_i (\beta_3^2 - \frac{1}{4}) - \beta_3(a_r^2 - a_i^2)\right\}\right], \label{eq18a} \\
			E_i &= \frac{\hbar^2}{2|m|^2} \left[m_r\left\{2 a_r a_i (\beta_3^2 - \frac{1}{4}) - \beta_3(a_r^2 - a_i^2)\right\} - m_i \left\{(a_r^2-a_i^2)(\beta_3^2 - \frac{1}{4}) + 2\beta_3 a_r a_i\right\}\right]. \label{eq18b}
		\end{align}
	\end{subequations}
	
	And the eigenfunction is given as, 
	
	\begin{equation}
		\psi(x) = exp\left[\left\{\frac{1}{2}+i\beta_3\right\}ax + i \beta_3 e^{-ax}\right]. \label{eq19}
	\end{equation}
	
	\section{Normalization of eigenfunctions} \label{sec4}
	
	The normalization of the eigenfunction is a fundamental requirement in quantum mechanics, arising directly from the probability postulate, which mandates that the total probability of finding a particle in the entire configuration space must equal unity. Mathematically, this ensures that the eigenfunction describes a physically admissible state.
	
	In conventional Hermitian quantum mechanics, this condition is straightforward; however, for non-Hermitian systems such as the present case of a complex mass diatomic molecule under action of complex Morse potential, the normalization process is more intricate and continues to be an active area of research. In this section, we investigate the normalization condition within the framework of extended complex phase space (ECPS) which is given by 
	
	\begin{equation}
		\int_{-\infty}^{\infty} \int_{-\infty}^{\infty} N^2 \psi^*\left(x_1,p_2\right) \psi\left(x_1,p_2\right) dx_1 dp_2 =N^2\int_{-\infty}^{\infty} \int_{-\infty}^{\infty} |\psi\left(x_1,p_2\right)|^2 dx_1 dp_2 = 1. \label{eq20}
	\end{equation}
	
	where $\psi(x_1,p_2)$ is the normalised eigenfunction of the system, and $\psi(x1,p2)*$ defines its complex conjugate. The normalised eigenfunction in terms of $\psi(x)$ and a normalisation constant, $N$ can be given from Eq. (\ref{eq19}) as, 
	
	\begin{equation}
		\psi\left(x\right) = N \exp\left[\left\{\frac{1}{2}+i\beta_3\right\}ax+i\beta_3e^{-ax}\right]. \label{eq21}
	\end{equation}
	
	Substituting Eq. (\ref{eq21}) into the normalization condition (\ref{eq20}), we have:
	
	\begin{equation}
		N^2\int_{-\infty}^{\infty} \int_{-\infty}^{\infty} |\psi\left(x_1,p_2\right)|^2 dx_1 dp_2 = N^2\int_{-\infty}^{\infty}\int_{-\infty}^{\infty}\exp\left[-2\left\{\alpha_1 x_1 + \beta_1 p_2\right\}\right] dx_1 dp_2 = 1. \label{eq22}
	\end{equation}
	
	where $\alpha_1$ and $\beta_1$ are real coefficients determined from the parameters of the system.
	
	The double integral in Eq. (\ref{eq22}) is found to diverge over the entire real domain due to the exponential term’s behavior. However, by considering the absolute values of $x_1$ and $p_2$, convergence is ensured, provided the both $\alpha_1$ and $\beta_1$ are positive real constants. Under this assumption, the integral can be separated into two one-dimensional integrals as:
		
	$$
	\int_{-\infty}^{\infty} \exp[-2\left\{\alpha_1|x_1|\right\}] dx_1 = \frac{1}{\alpha_1}, \quad 	\int_{-\infty}^{\infty} \exp[-2\left\{\beta_1|p_2|\right\}] dp_2 = \frac{1}{\beta_1};\quad \alpha_1>0, \beta_1>0 
	$$
	
	Thus, the two-dimensional integral in Eq. (\ref{eq22}) reduces to:
	
	\begin{equation}
		N^2 \int_{-\infty}^{\infty} \int_{-\infty}^{\infty} |\psi\left(x_1,p_2\right)|^2 dx_1 dp_2 = N^2 \frac{1}{\alpha_1\beta_1} = 1. \label{eq23}
	\end{equation}
	
	Solving for $N$, the normalization constant is obtained as:
	
	\begin{equation}
		N = \sqrt{\alpha_1 \beta_1}. \label{eq24}
	\end{equation}
	
	For the parameters $\alpha_1$ and $\beta_1$ to remain positive, the following constraints on $\beta_3$ must be satisfied:
	
	\begin{subequations} \label{eq25}
		\begin{align}
			\beta_3 &> -\frac{a_i}{2 a_r}, \label{eq25a} \\
			\beta_3 &> \frac{a_r}{2 a_i}. \label{eq25b}
		\end{align}
	\end{subequations}
	
	where $a_r$ and $a_i$ denote the real and imaginary components of the complex Morse parameter $a$, respectively. 
	
	Thus, the normalized eigenfunction and probability density are expressed as:
	
	\begin{subequations} \label{eq26}
		\begin{align}
			\psi(x_1,p_2) = \sqrt{\alpha_1 \beta_1} \exp{\left[\left\{\frac{1}{2} + i \beta_3\right\}ax + i\beta_3 e^{-ax}\right]}, \label{eq26a} \\
			|\psi(x_1, p_2)|^2 = \alpha_1 \beta_1 \exp{\left[-2\left\{\alpha_1|x_1| + \beta_1 |p_2|\right\}\right]}. \label{eq26b}
		\end{align}
	\end{subequations}
	
The total eienfunction can take the form

	\begin{equation}
		\psi(x_1,p_2) = \sqrt{\alpha_1 \beta_1} \exp{\left[\left\{\frac{1}{2} + i \beta_3\right\}ax + i\beta_3 e^{-ax} - \frac{E_i t}{\hbar}\right]} \exp{\left(i \frac{E_r t}{\hbar}\right)}. \label{eq27}
	\end{equation}
	
	It is implicit that $E_r$ which real, measurable energy of the system represents the oscillatory phase of the eigenfunction that governs the temporal phase evolution of the quantum state and also contributes to the quantum coherence and wave-like interference properties of the state. Physically, $E_r$ is the part responsible for observable motion, resonance frequency, or localization of the eigenfunction in potential wells. On the other hand, $E_i$ is responsible for damping or amplification of probability density thus controlling the amplitude modulation of the wavefunction over time. Thus, $E_i$ dictates whether the state is stable ($E_i = 0$), metastable (small $E_i$, $E_i \approx 0$) characterized by slow decay, long-lived resonance or Unstable (large $E_i > 0$ or $< 0$) implying rapid decay or amplification. Thus a small positive $E_i$ corresponds to weakly decaying, quasi-stable states, similar to resonant or metastable matter–like states that persist but do not radiate significantly. On the other hand, a large $E_i$ indicates rapid dissipation or instability, implying non-physical or transient quantum states. A zero $E_i$ corresponds to non-radiative, purely real-energy states.
	
	\section{Analysis of Eigenfunctions and eigenvalues- General case} \label{sec5}
	
	\begin{figure}[h!]
		\centering
	
		\begin{subfigure}[b]{0.45\textwidth}
			\centering
			\includegraphics[width=\textwidth]{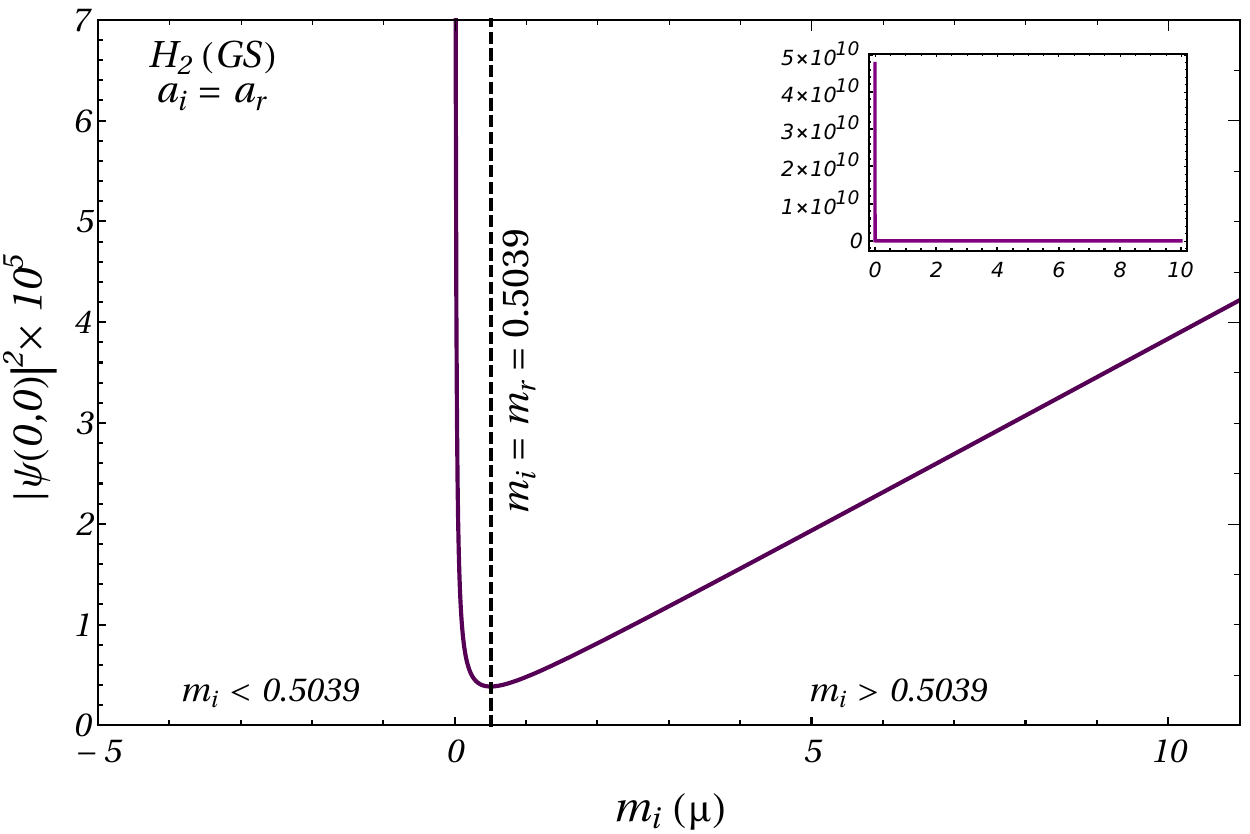}
			\caption{}
			\label{fig1a}
		\end{subfigure}
		\hfill
		\begin{subfigure}[b]{0.45\textwidth}
			\centering
			\includegraphics[width=\textwidth]{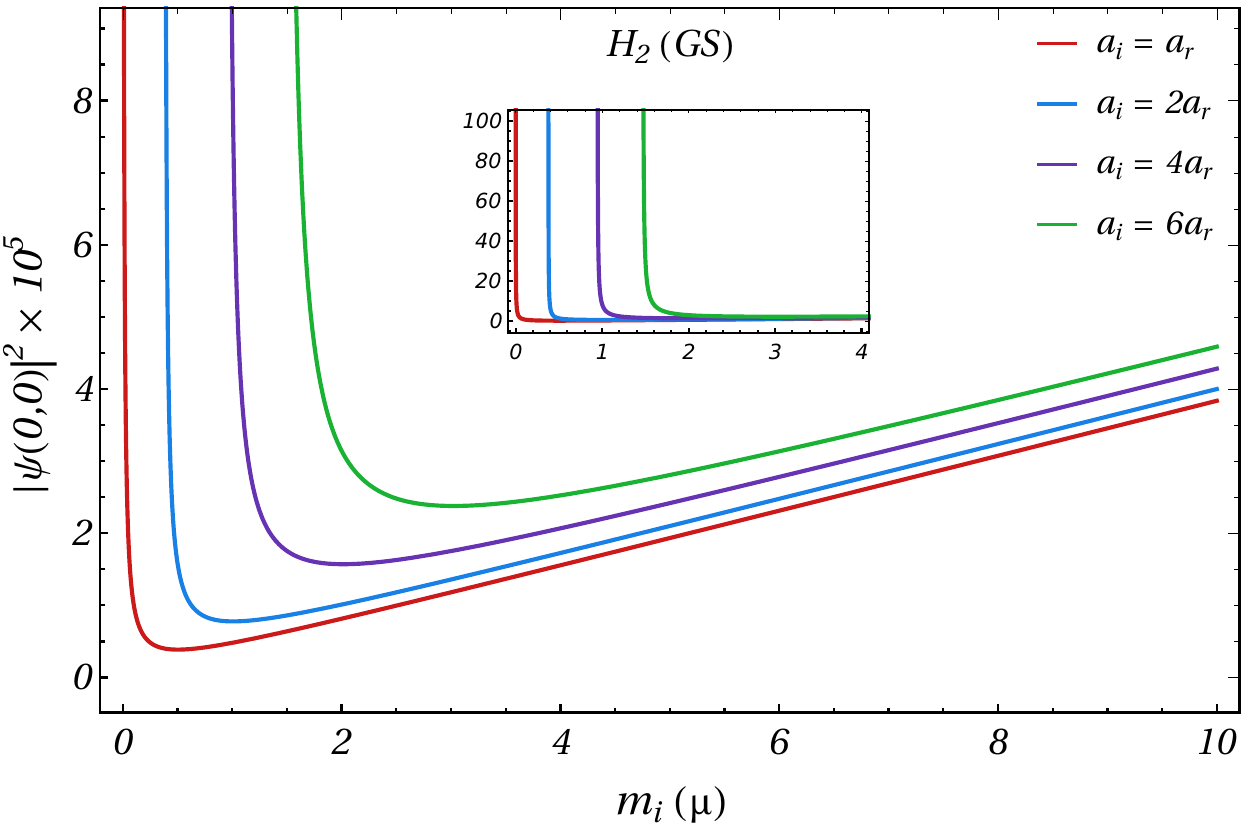}
			\caption{}
			\label{fig1b}
		\end{subfigure}
		
		\vspace{0.5cm}
		
		\begin{subfigure}[b]{0.45\textwidth}
			\centering
			\includegraphics[width=\textwidth]{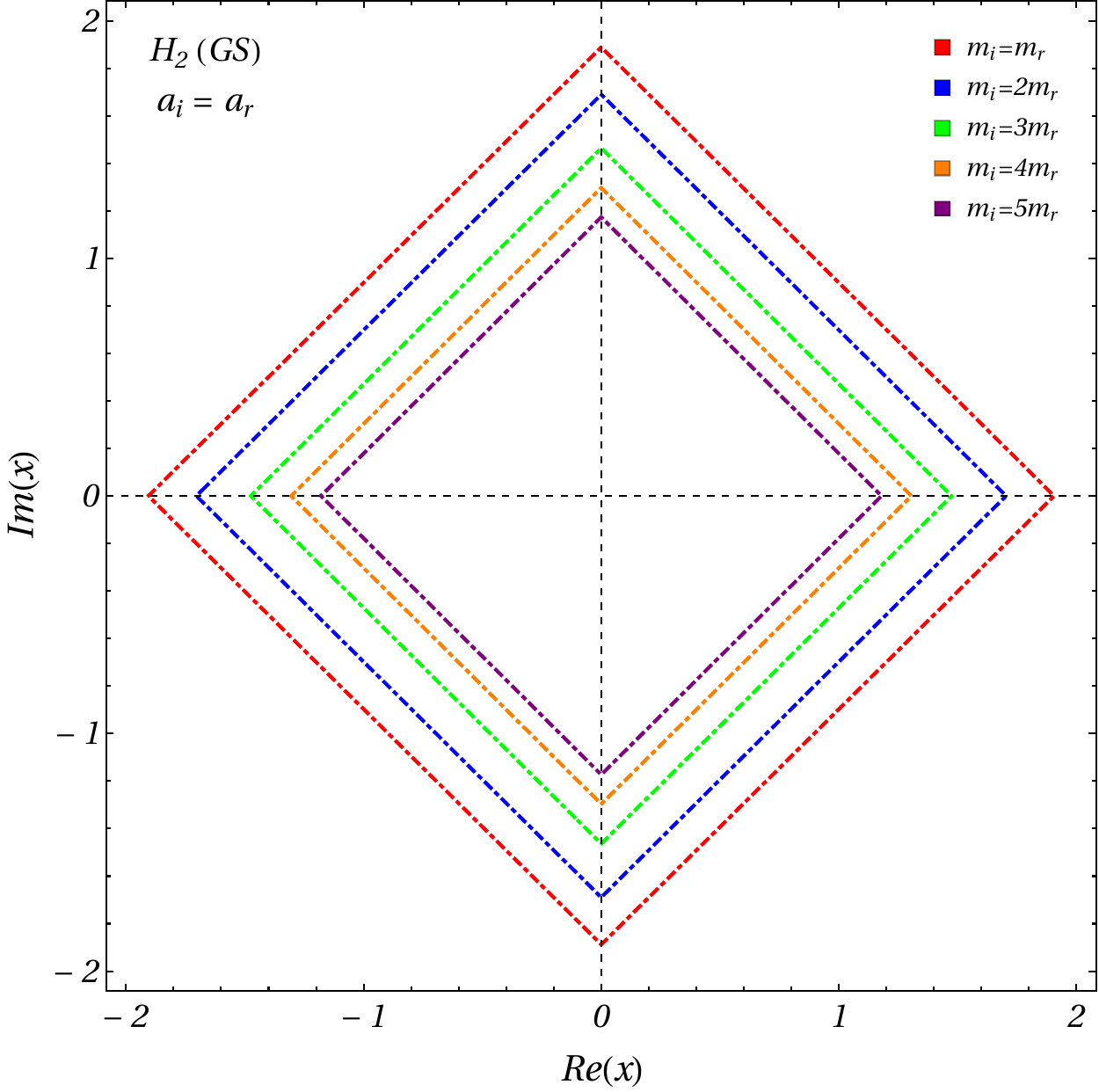}
			\caption{}
			\label{fig1c}
		\end{subfigure}
		\hfill
		\begin{subfigure}[b]{0.45\textwidth}
			\centering
			\includegraphics[width=\textwidth]{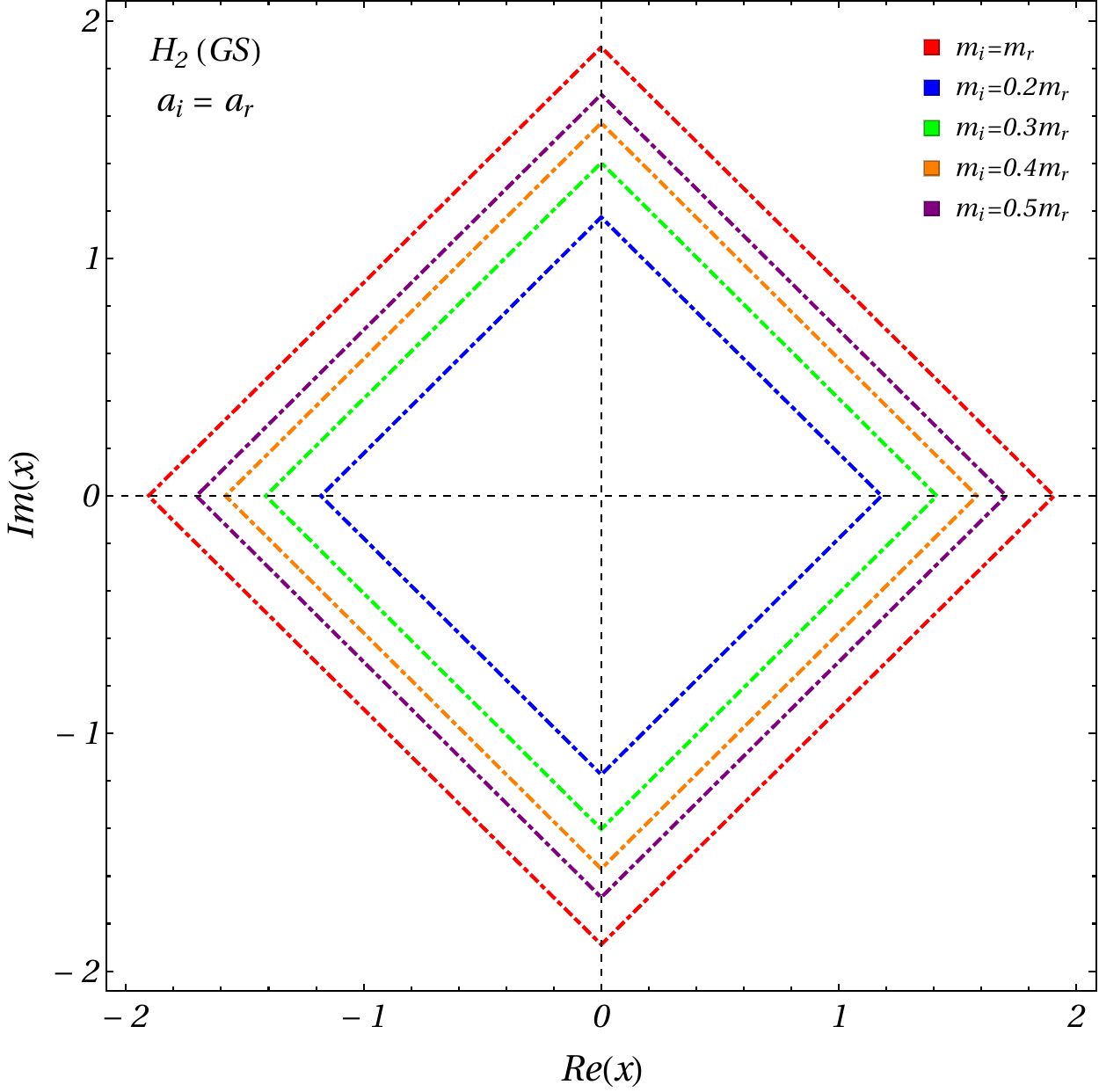}
			\caption{}
			\label{fig1d}
		\end{subfigure}
		
		\caption{The ground state (GS) probability density plots of the hydrogen ($H_2$) molecule with respect to $m_i$. (a) shows variation of peak value of the probability density with respect to $m_i$ for the fixed value of $a_i$ taken as $a_i = a_r$. Here, the minima exists at $m_i = m_r$. (b) shows dependency of peak probability density on complex mass parameter for various values of $a_i$. (c) and (d) reveals the variation of spartial confingement of probability density with varying $m_i$ and fixed $a_i$ as $a_i = a_r$. The value of real parameters for the $H_2$ molecule are taken as $V_{or} = 38266$ $cm^{-1}$, $a_r = 1.868$ $\AA^{-1}$, $m_r = 0.5039$ $\mu$.}
		\label{fig1}
	\end{figure}
	
	\begin{figure}[h!]
		\centering
		
		\begin{subfigure}[b]{0.45\textwidth}
			\centering
			\includegraphics[width=\textwidth]{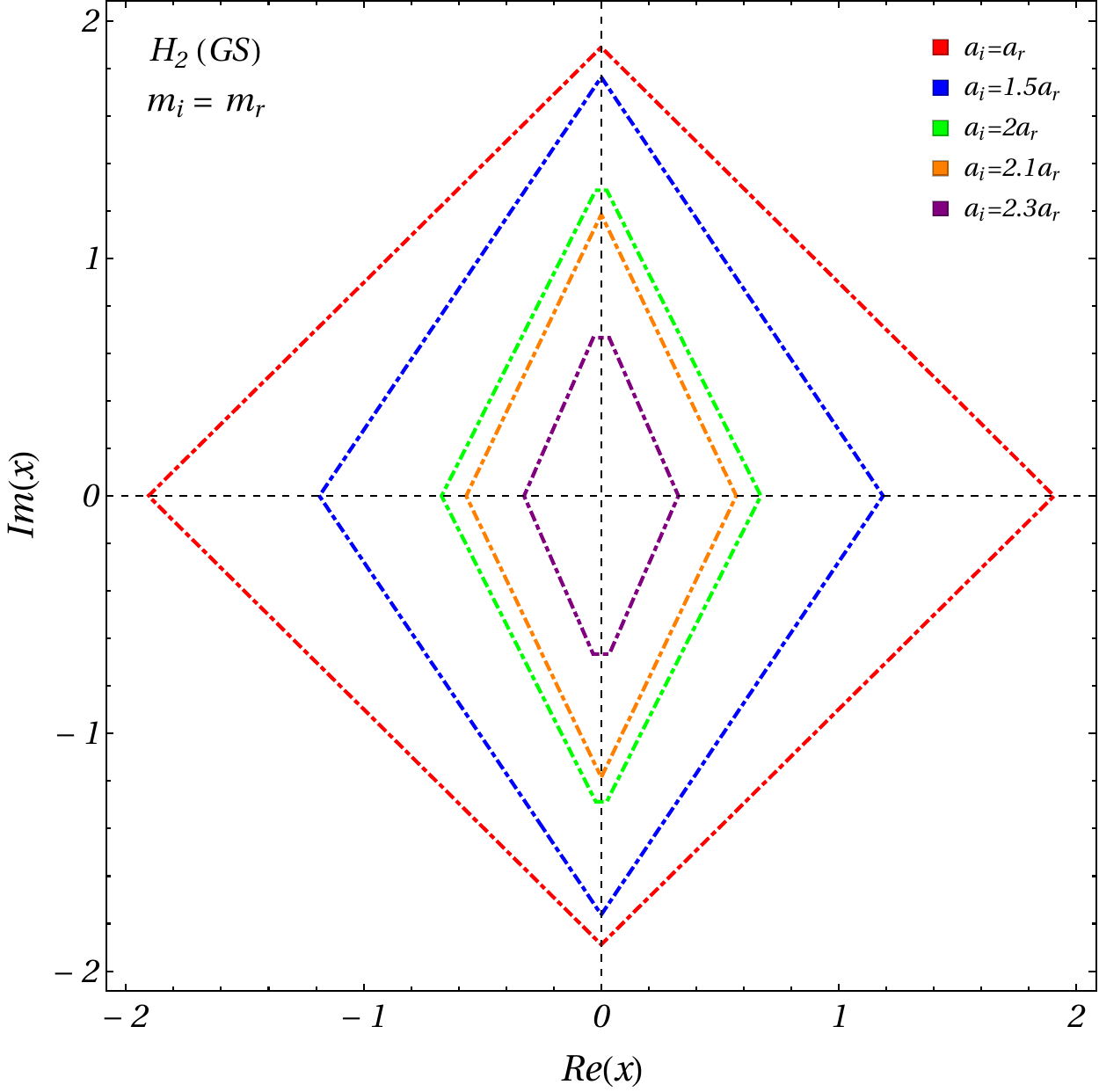}
			\caption{}
			\label{fig2a}
		\end{subfigure}
		\hfill
		\begin{subfigure}[b]{0.45\textwidth}
			\centering
			\includegraphics[width=\textwidth]{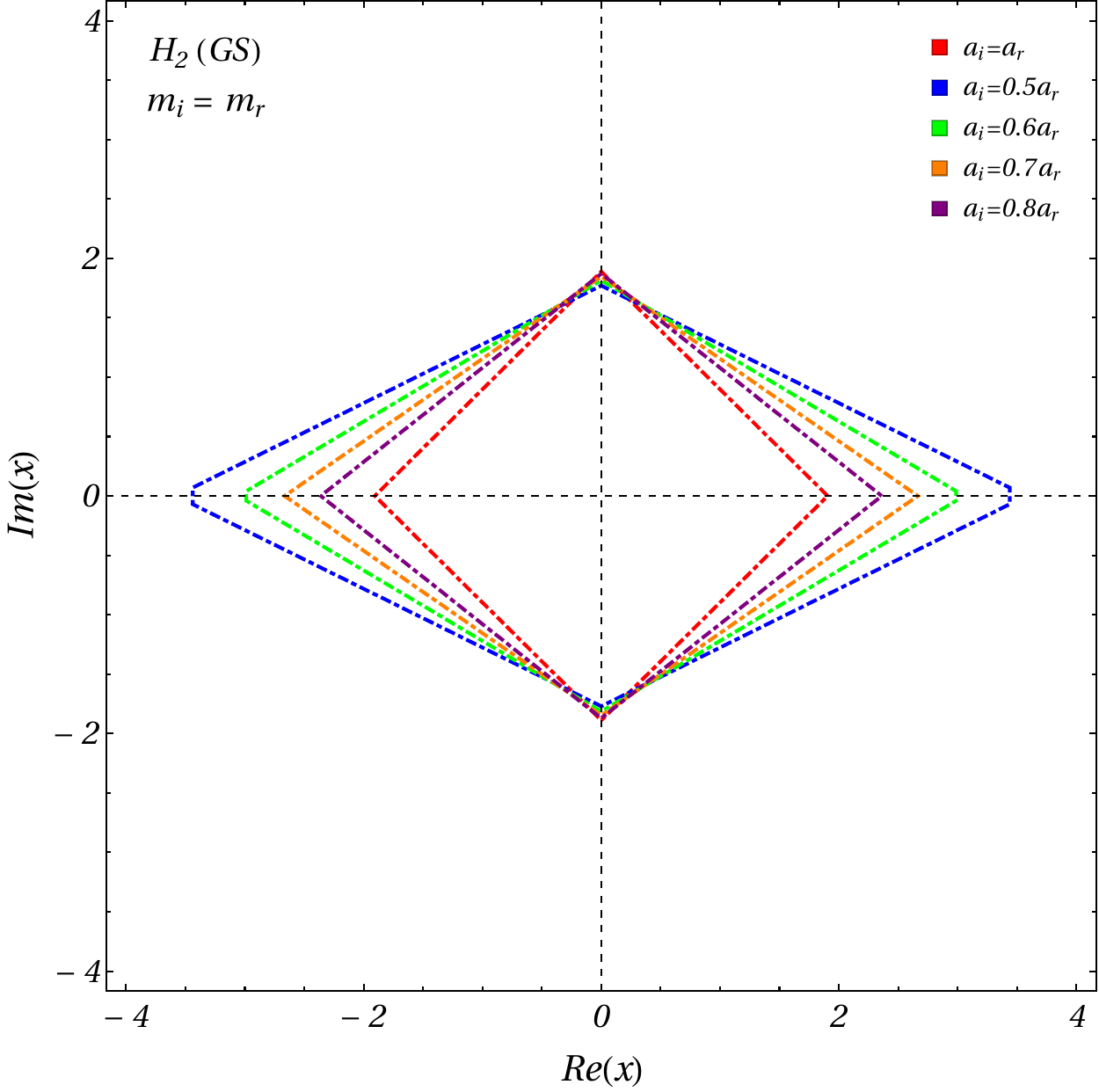}
			\caption{}
			\label{fig2b}
		\end{subfigure}
		
		\vspace{0.5cm}
		
		\begin{subfigure}[b]{0.45\textwidth}
			\centering
			\includegraphics[width=\textwidth]{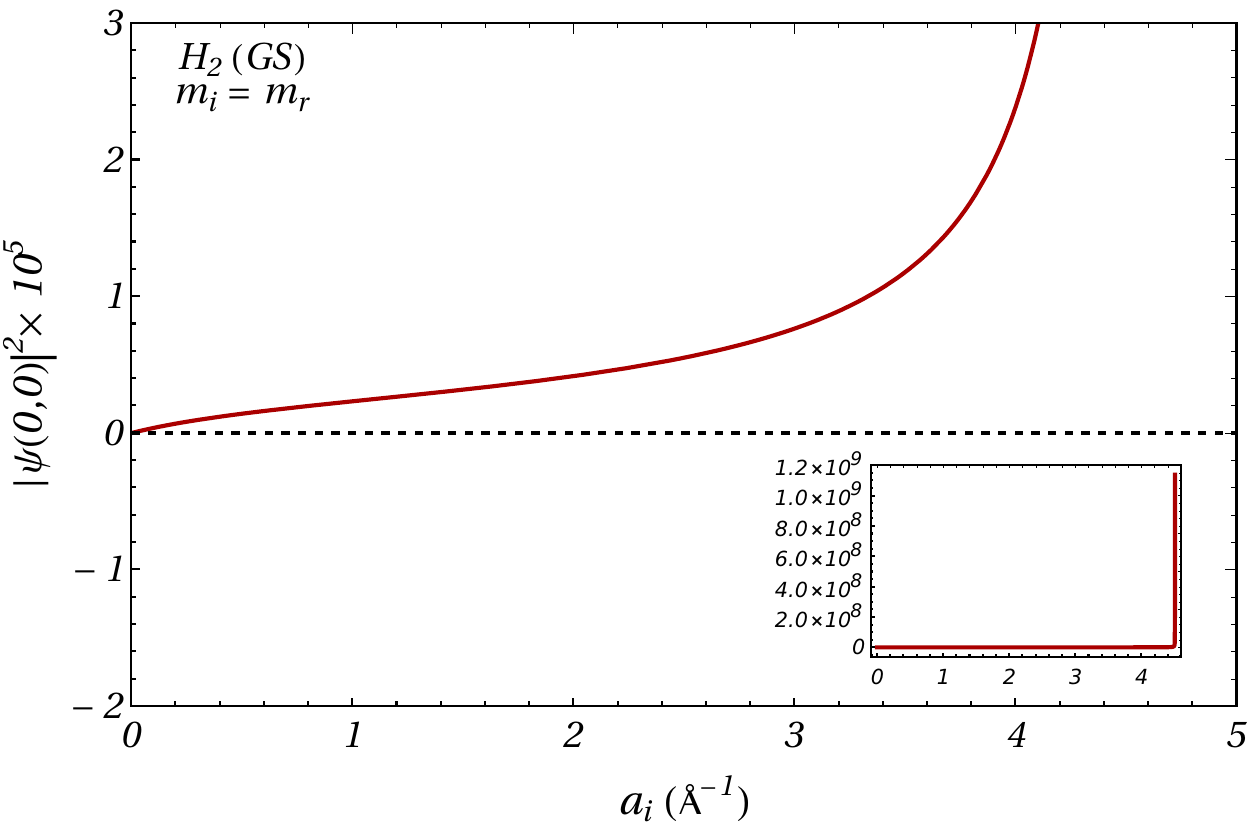}
			\caption{}
			\label{fig2c}
		\end{subfigure}
		\hfill
		\begin{subfigure}[b]{0.45\textwidth}
			\centering
			\includegraphics[width=\textwidth]{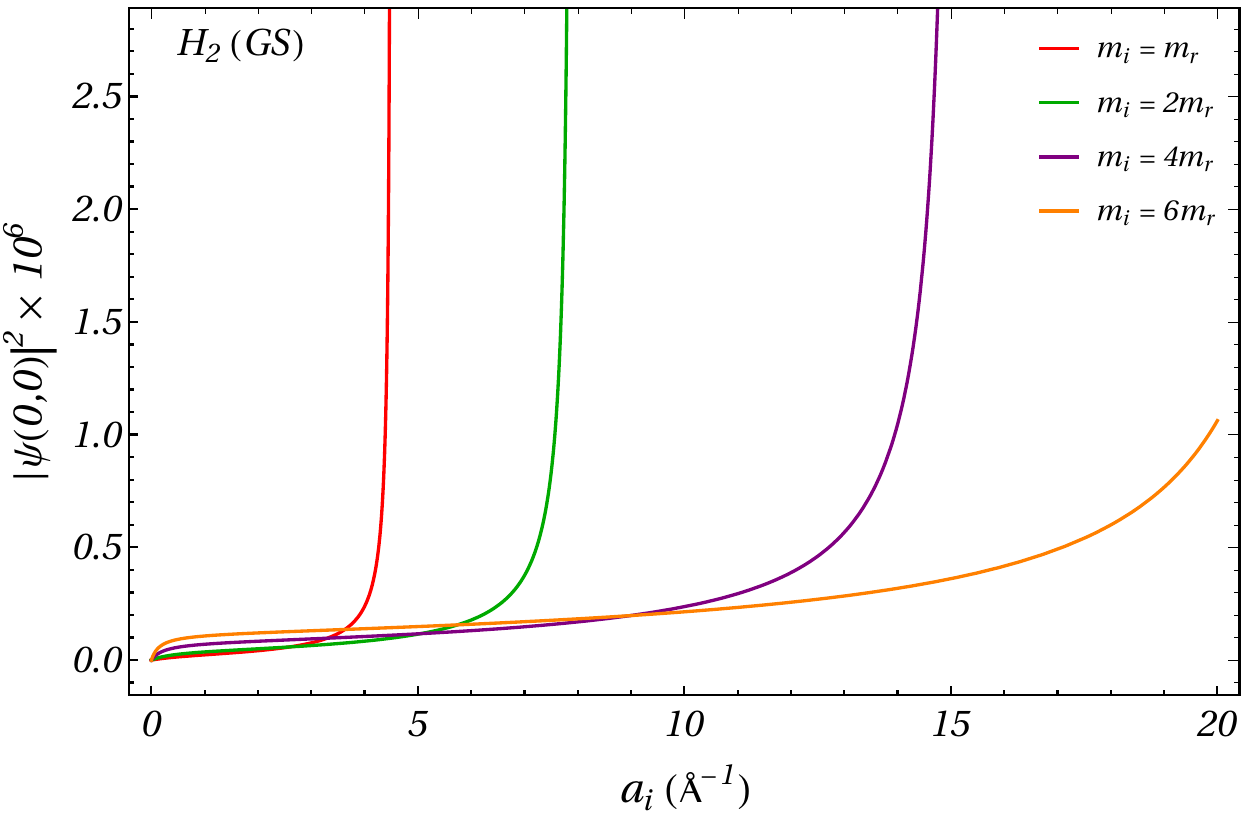}
			\caption{}
			\label{fig2d}
		\end{subfigure}
		
		\caption{The ground state (GS) probability density plots of the hydrogen ($H_2$) molecule with respect to $a_i$. (a) and (b) reveals the variation of spartial confingement of probability density with varying $a_i$ and fixed value of $m_i$ taken as $m_i = m_r$. (c) shows variation of peak value of the probability density with respect to $a_i$ and fixed value of $m_i$, as $m_i = m_r$. (d) shows dependency of peak probability density on complex Morse parameter for various values of $m_i$. The value of real parameters for the $H_2$ molecule are taken as $V_{or} = 38266$ $cm^{-1}$, $a_r = 1.868$ $\AA^{-1}$, $m_r = 0.5039$ $\mu$.}
		\label{fig2}
	\end{figure}
	
	\begin{figure}[h!]
		\centering
		
		\begin{subfigure}[b]{0.45\textwidth}
			\centering
			\includegraphics[width=\textwidth]{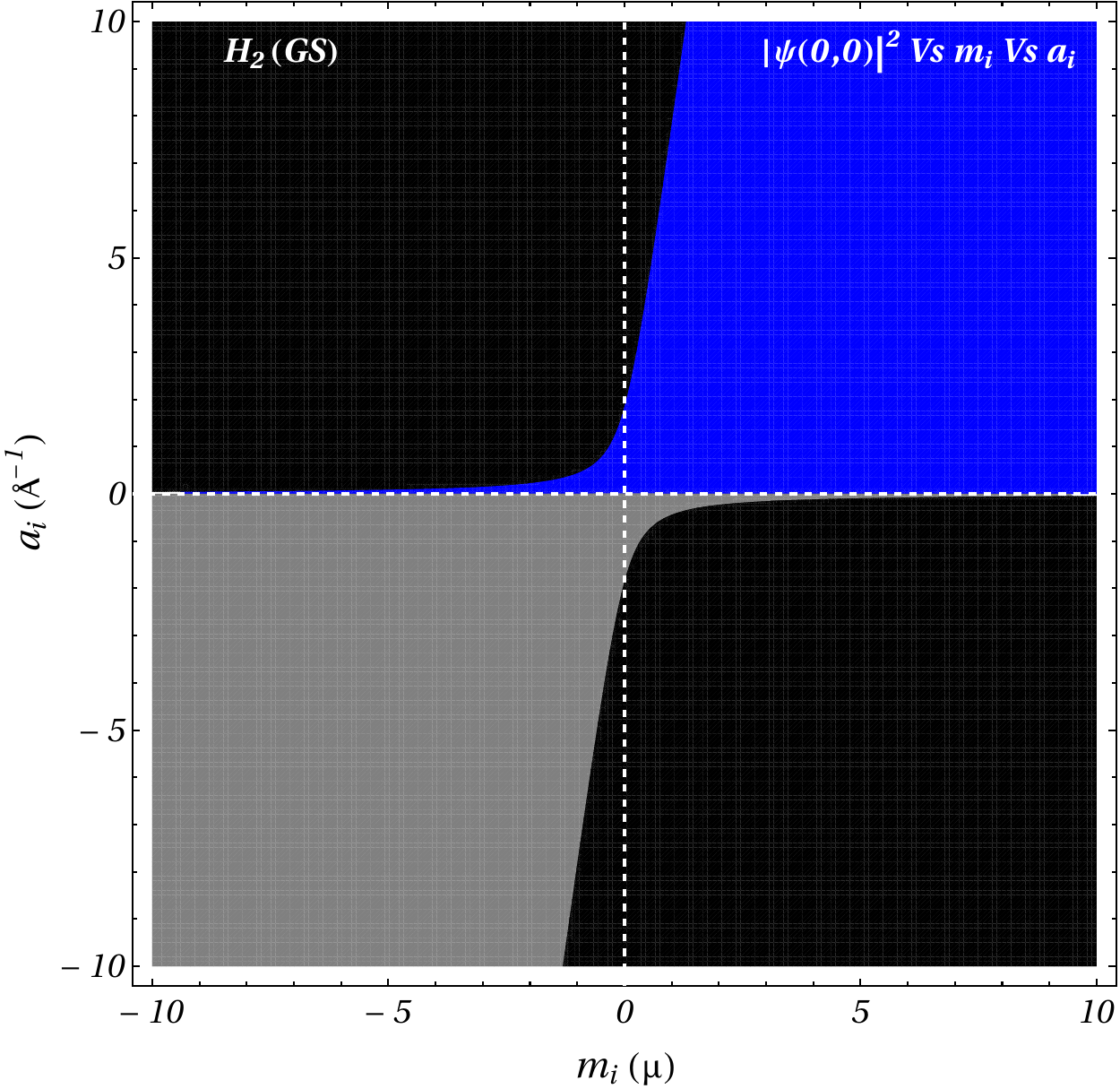}
			\caption{}
			\label{fig3a}
		\end{subfigure}
		\hfill
		\begin{subfigure}[b]{0.47\textwidth}
			\centering
			\includegraphics[width=\textwidth]{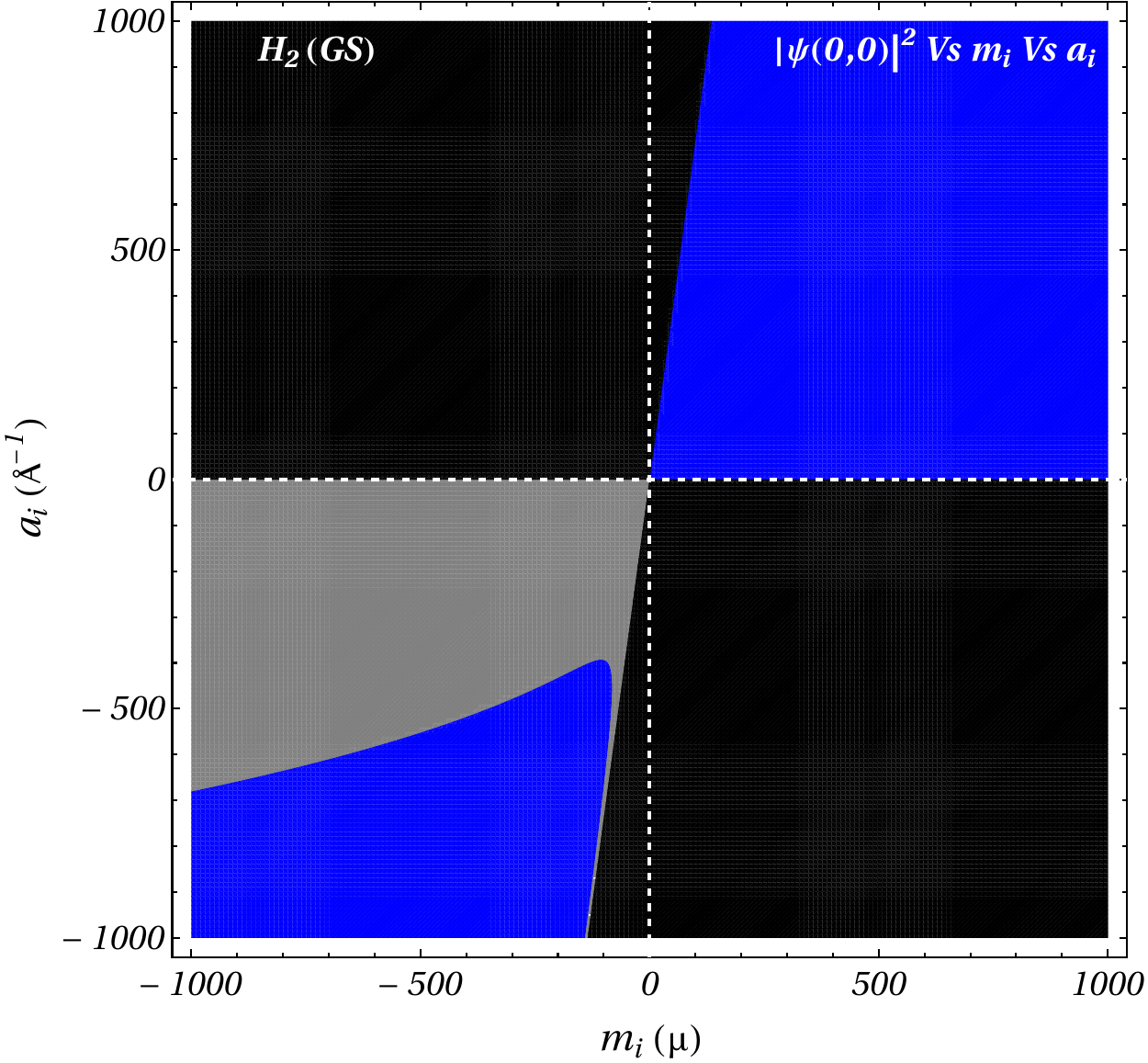}
			\caption{}
			\label{fig3b}
		\end{subfigure}
		
		\caption{The ground state (GS) peak probability density (PPD) region plots of the hydrogen ($H_2$) molecule in the parametric space defined by ($a_i$, $m_i$). The blue shaded region shows permissible values of ($a_i$, $m_i$) for positive PPD and region shaded with black and grey corresponds to the values of ($a_i$, $m_i$) for which PPD is non-normalizable and negative, respectively. Here, The value of real parameters for the $H_2$ molecule are taken as $V_{or} = 38266$ $cm^{-1}$, $a_r = 1.868$ $\AA^{-1}$, $m_r = 0.5039$ $\mu$.}
		\label{fig3}
	\end{figure}
	
	\begin{figure}[h!]
		\centering
		
		\begin{subfigure}[b]{0.45\textwidth}
			\centering
			\includegraphics[width=\textwidth]{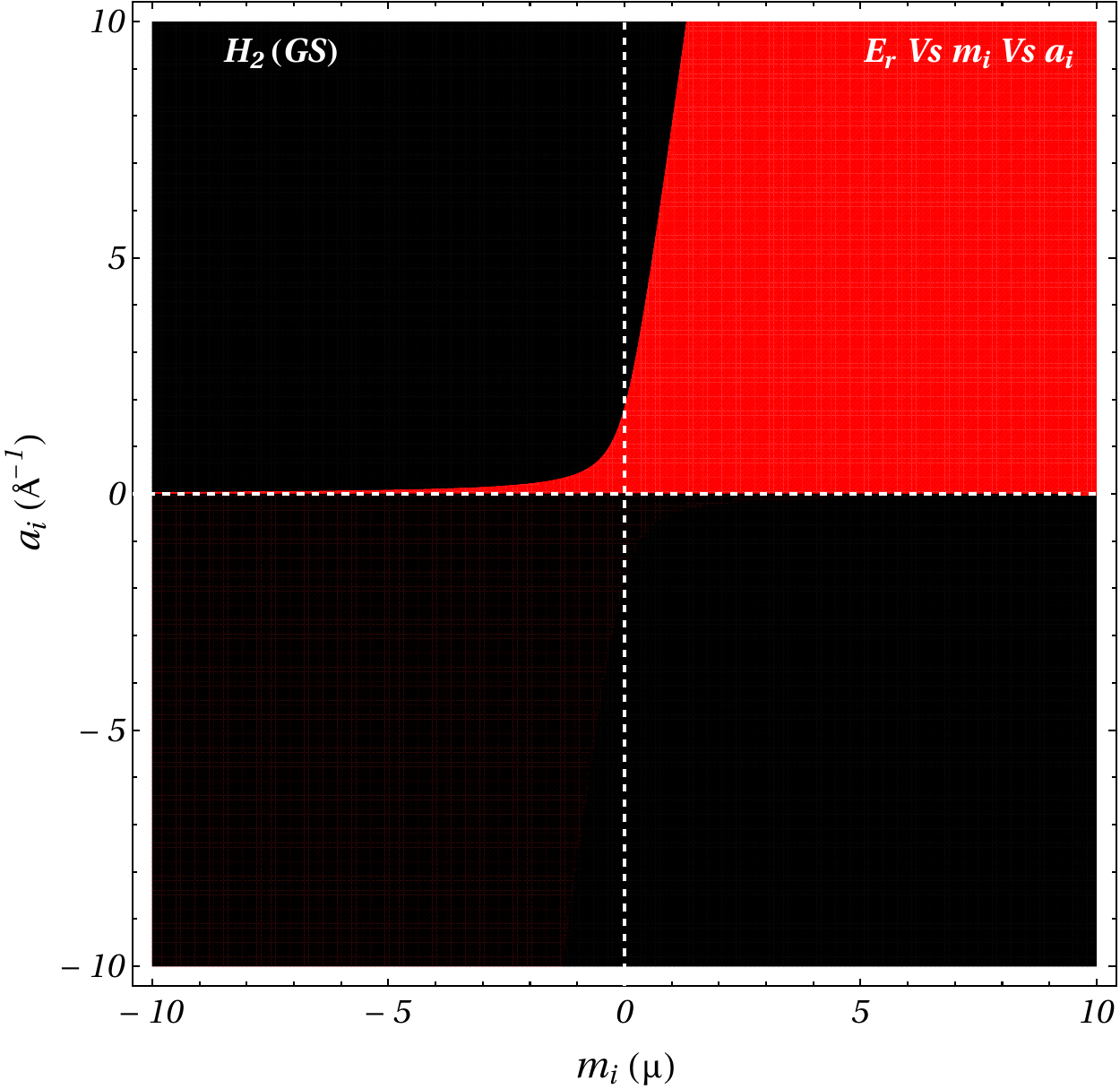}
			\caption{}
			\label{fig4a}
		\end{subfigure}
		\hfill
		\begin{subfigure}[b]{0.47\textwidth}
			\centering
			\includegraphics[width=\textwidth]{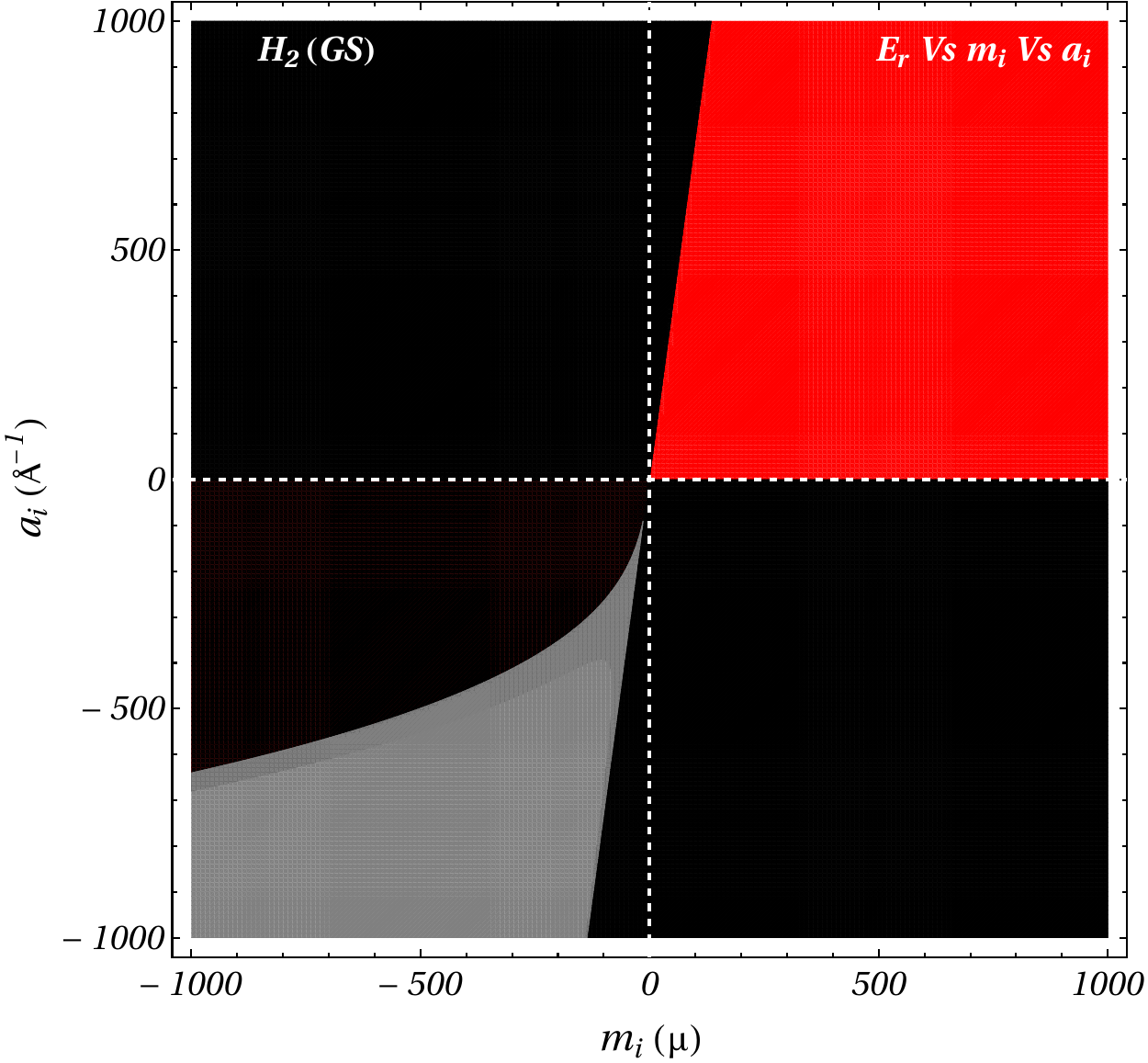}
			\caption{}
			\label{fig4b}
		\end{subfigure}
		
		\vspace{0.5cm}
		
		\begin{subfigure}[b]{0.465\textwidth}
			\centering
			\includegraphics[width=\textwidth]{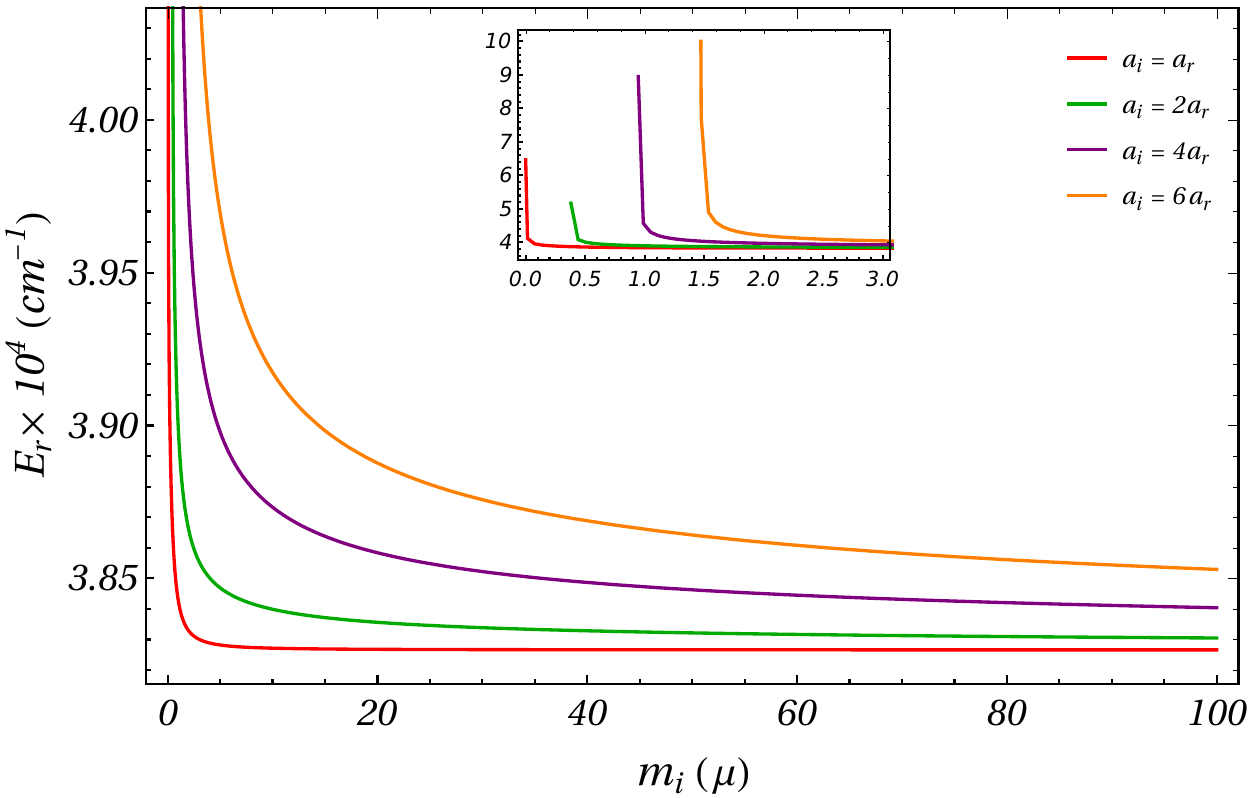}
			\caption{}
			\label{fig4c}
		\end{subfigure}
		\hfill
		\begin{subfigure}[b]{0.45\textwidth}
			\centering
			\includegraphics[width=\textwidth]{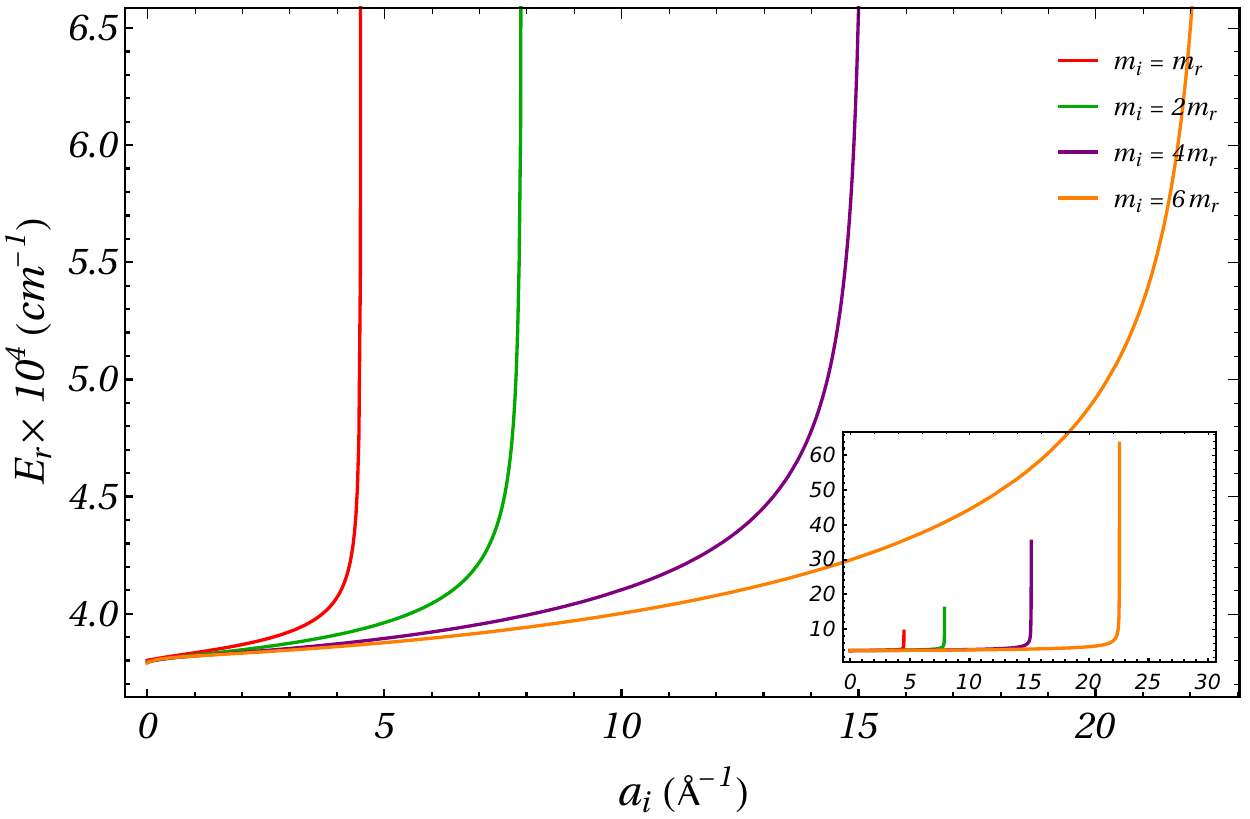}
			\caption{}
			\label{fig4d}
		\end{subfigure}
		
		\caption{The Variation of the real part of the ground state (GS) energy eigenfunction ($E_r$) of the hydrogen molecule ($H_2$) with respect to imaginary parameters, ($a_i$, $m_i$). (a) and (b) reveals region plot in the parameteric space of ($a_i$, $m_i$). The region shaded with red and gray corresponds to positive and negative value of $E_r$, i.e, defines permissible region of ($a_i$, $m_i$), whereas the black shaded region corresponds to region where PPD is non-normalizable or negative. (c) reveals the nature of $E_r$ with respect to $m_i$ for various values of $a_i$ and (d) shows the nature of $E_r$ with respect to $a_i$ for various values of $m_i$. Here, The value of real parameters for the $H_2$ molecule are taken as $V_{or} = 38266$ $cm^{-1}$, $a_r = 1.868$ $\AA^{-1}$, $m_r = 0.5039$ $\mu$.}
		\label{fig4}
	\end{figure}
	
	\begin{figure}[h!]
		\centering
		
		\begin{subfigure}[b]{0.45\textwidth}
			\centering
			\includegraphics[width=\textwidth]{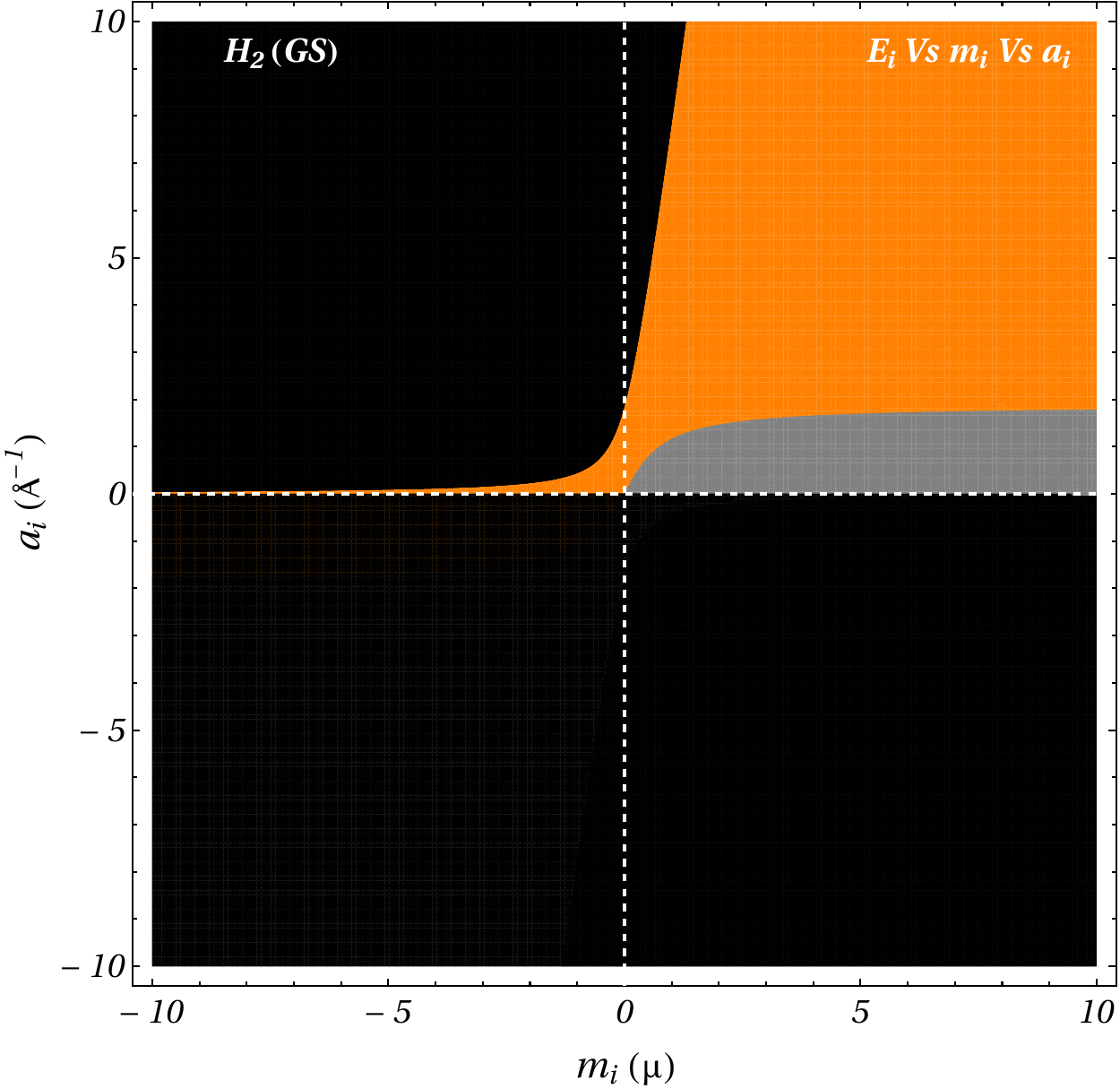}
			\caption{}
			\label{fig5a}
		\end{subfigure}
		\hfill
		\begin{subfigure}[b]{0.47\textwidth}
			\centering
			\includegraphics[width=\textwidth]{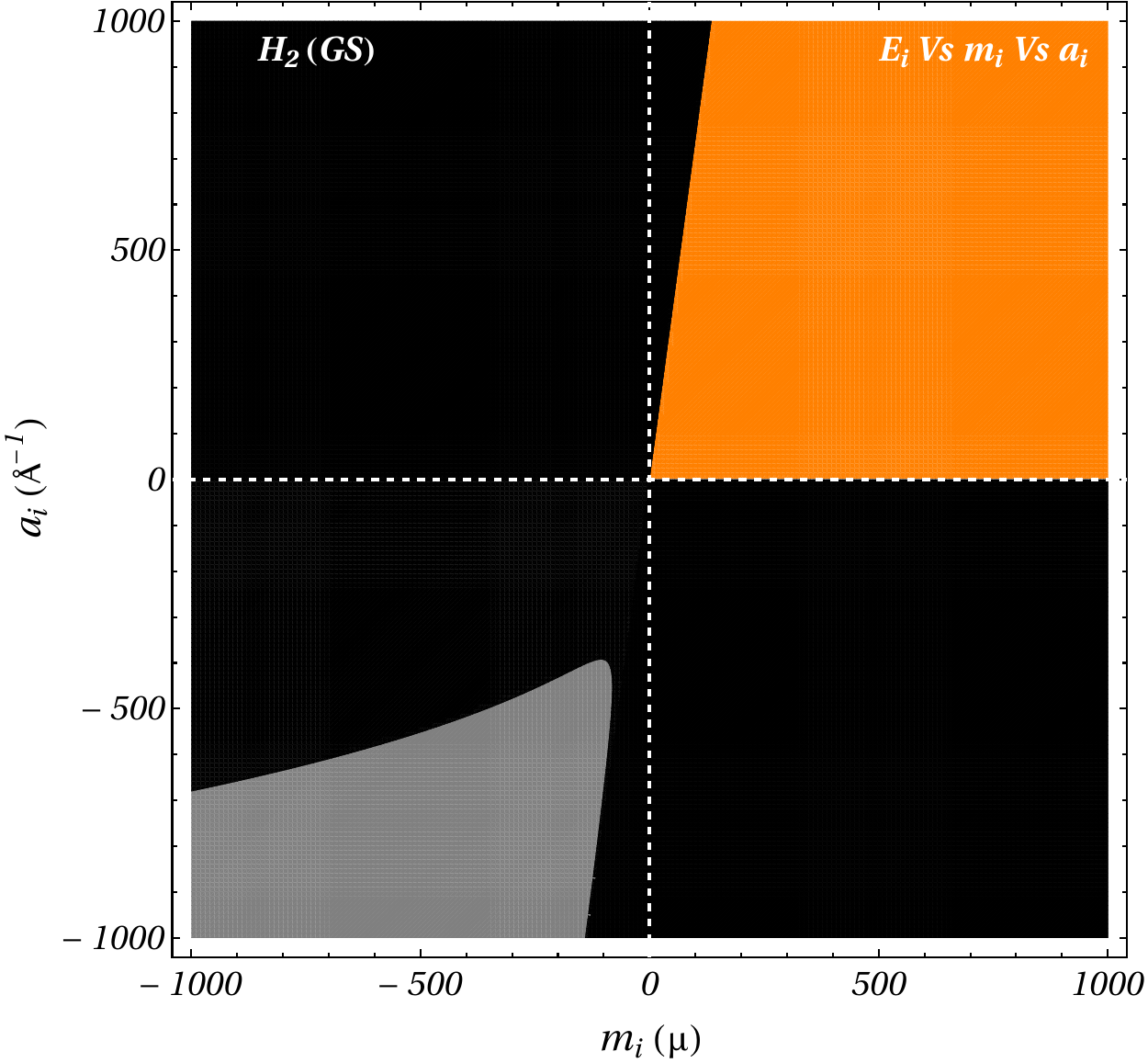}
			\caption{}
			\label{fig5b}
		\end{subfigure}
		
		\vspace{0.5cm}
		
		\begin{subfigure}[b]{0.465\textwidth}
			\centering
			\includegraphics[width=\textwidth]{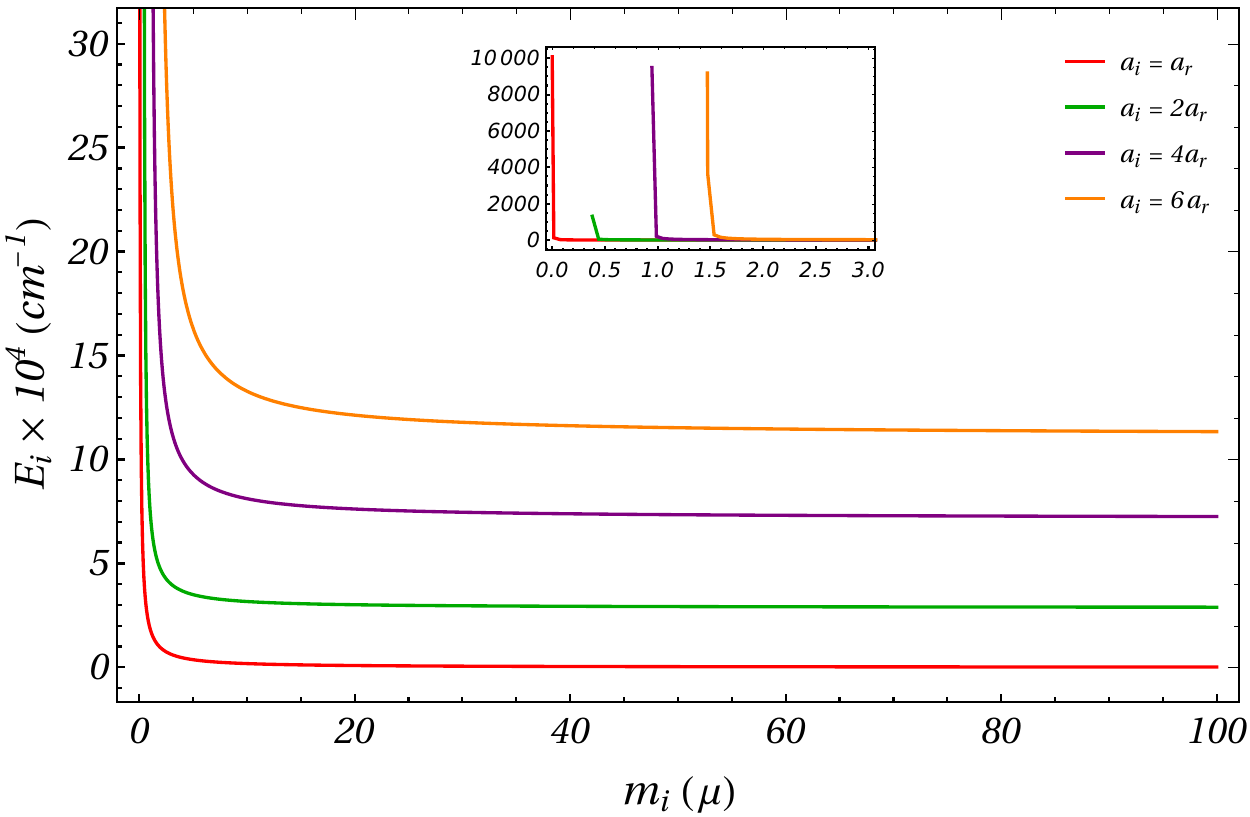}
			\caption{}
			\label{fig5c}
		\end{subfigure}
		\hfill
		\begin{subfigure}[b]{0.45\textwidth}
			\centering
			\includegraphics[width=\textwidth]{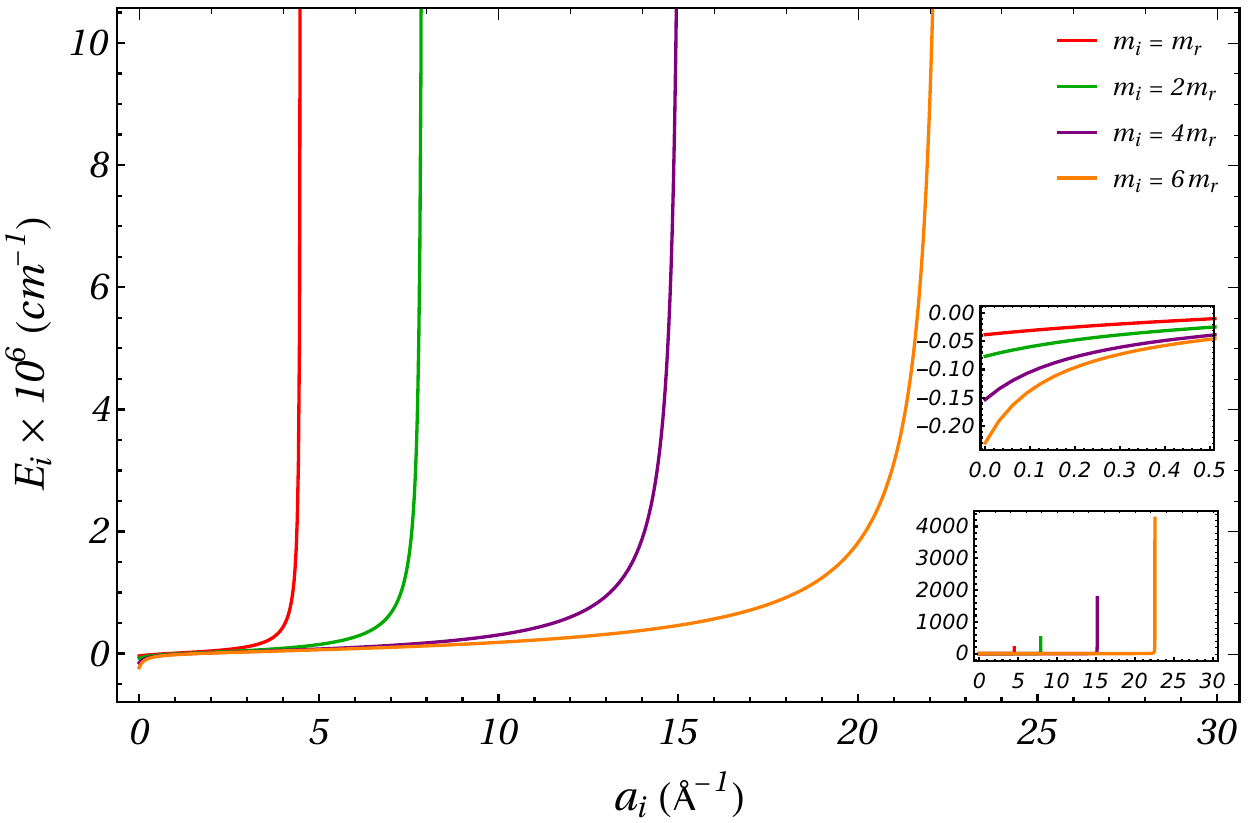}
			\caption{}
			\label{fig5d}
		\end{subfigure}
		
		\caption{The Variation of the Imaginary part of the ground state (GS) energy eigenfunction ($E_i$) of the hydrogen ($H_2$) molecule with respect to imaginary parameters, ($a_i$, $m_i$). (a) and (b) reveals region plot in the parameteric space of ($a_i$, $m_i$). The region shaded with orange and gray corresponds to positive and negative value of $E_i$, i.e, defines permissible region of ($a_i$, $m_i$), whereas the black shaded region corresponds to region where PPD is non-normalizable or negative. (c) reveals the nature of $E_i$ with respect to $m_i$ for various values of $a_i$ and (d) shows the nature of $E_i$ with respect to $a_i$ for various values of $m_i$. Here, The value of real parameters for the $H_2$ molecule are taken as $V_{or} = 38266$ $cm^{-1}$, $a_r = 1.868$ $\AA^{-1}$, $m_r = 0.5039$ $\mu$.}
		\label{fig5}
	\end{figure}

	To understand the quantum behaviour of a diatomic molecule with complex mass under the influence of a complex Morse potential, it is essential to graphically analyze its probability density distribution and the corresponding energy spectrum in the extended complex phase space.
	
	These graphical analyses reveal how variations in the imaginary components of the mass ($m_i$) and Morse parameter ($a_i$) affect the spatial probability distribution, peak probability density (PPD), and real and imaginary parts of the energy eigenvalues. 
	
	Such studies help identify critical parameter domains beyond which the system transitions from solvable quantum system with normalized eigenfunction and finite eigenvalues to chaotic  behaviour, thereby establishing bounds of physical admissibility.
	
	In this section, we discuss the findings arising from Plots \ref{fig1}-\ref{fig5}, which provide visual insights into the probability density, PPD, and energy spectrum as functions of the complex parameters $m_i$ and $a_i$.
	
	\subsection{Probability Density Variation with Imaginary Mass ($m_i$)} \label{subsec5.1}
	
	Plot \ref{fig1} summarizes how peak probability density in the hydrogen molecule evolves across weak, intermediate, and strong imaginary-mass regimes. Figure \ref{fig1a} illustrates the dependence of the ground-state probability density at the origin on the imaginary mass parameter $m_i$ under the parametric condition $a_i =  a_r$. The curve displays an initial rapid decline for small $m_i$ till  the minimum value of the function is attained when $m_i = m_r$. This indicate that even weak complex mass contributions reduce central localization of the ground-state eigenfunction. Beyond the minimum, the probability density  increases steadily and approaches a near-linear growth for larger values of $m_i$. This non-monotonic behavior demonstrates that while small imaginary mass tends to delocalize the eigenfunction, stronger complex mass components enhance localization by reshaping the effective potential landscape. The inset further confirms that at very large $m_i$, the probability density grows dramatically, signalling a regime in which complex mass dominates the spatial behavior of the eigenfunction. This leads to a physical condition where the low imaginary mass leads to high localization at the origin, implying the eigenfunction is sharply peaked.
	
	The results of the plot \ref{fig1a} is explored for complex Hydrogen-like molecules with varying ai in  plot \ref{fig1b} that illustrates the variation of the ground-state peak probability density $|\psi(0,0)|^2$ as a function of the complex mass parameter $m_i$ for different initial mass scales $a_i$. For all values of $a_i$, the probability density exhibits a sharp increase as $m_i$ tends to its lowest possible critical value $m_c$ $(m_i \rightarrow m_c)$, followed by a pronounced minimum before rising gradually in the higher-mass regime. This non-monotonic behaviour reflects the strong sensitivity of the ground-state localization to the effective mass near the origin. Larger $a_i$ values shift the minimum toward higher masses and enhance the magnitude of $|\psi(0,0)|^2$ indicating stronger eigenfunction compression under broader mass profiles. Overall, the curves highlight how modified mass distributions significantly influence the spatial concentration of the ground-state eigenfunction.
	
	For the condition in \ref{fig1a}, where $a_i = a_r$, Figure \ref{fig1c} illustrates the contour of the ground-state probability density of the hydrogen molecule plotted over the complex -plane for several values of the imaginary mass parameter $m_i$ is greater than mr. The contours exhibit a characteristic diamond-shaped structure, symmetric about both real and imaginary axes, reflecting the inherent symmetry of the underlying complex mass framework. As $m_i$ increases, the contours shrink inward, indicating that the spatial extent of the probability distribution diminish in the complex plane. This demonstrates that larger imaginary mass components enhances the localization strength of the ground state, effectively diminishing the distribution while preserving the geometric symmetry of the contour pattern. 
	
	Figure \ref{fig1d} presents the corresponding contour plots of the ground-state probability density under  the prescription $a_i = a_r$ where the imaginary mass is less than of the real mass, $m_i < m_r$. The contours retain the same diamond-like topology seen in the first panel,. As the value of mi decreases, the area under the contour shrinks, thus pointing towards the delocalization effect induced by the imaginary mass. This suggest that the spectral and spatial properties of the non-Hermitian hydrogen molecule depend sensitively not only on the magnitude of $m_i$ but also on how it is parametrized.

	\subsection{Probability Density Variation with Imaginary Morse Parameter ($a_i$)} \label{subsec5.2}
	
	Plot \ref{fig2} depicts the dependence of the probability density function on the imaginary part of the Morse parameter ($a_i$) for $m_i = m_r$, with other parameters held fixed. Figures \ref{fig2a} and \ref{fig2b} show that the area of existence monotonically decreases as $a_i$ increase. Figure \ref{fig2c} demonstrates that the PPD grows rapidly as $a_i$ approches $4.5\AA$, implying the existence of an upper bound for $a_i$. Beyond this bound, the eigenfunction becomes highly localized, and the system transitions to a non-quantum regime.This establishes a critical value $a_c$ beyond which the normalization condition fails, and the solution ceases to represent a physical quantum state.
	
	The plot \ref{fig2d} presents the dependence of the peak probability density $|\psi(0,0)|^2$ on the Morse parameter $a_i$ for several fixed values of the position-dependent mass $m_i$. For each $m_i$, the probability density initially increases slowly with $a_i$. As $a_i$ grows, however, $|\psi(0,0)|^2$ rises sharply, eventually diverging near a critical value that shifts to higher $a_i$ for larger $m_i$. These trends emphasize the nonlinear interplay between Morse parameter and effective mass in determining ground-state locatization. 
	
	\subsection{Critical Domains for $m_i$ and $a_i$: Existence and Quantum Admissibility} \label{subsec5.3}
	
	Plot \ref{fig3} depict a 3D parameter space projection for the Hydrogen molecule ($H_2$), showing the PPD as a function of imaginary mass component $m_i$ and imaginary potential range $a_i$.
	
	This study investigates a complex mass Schrödinger system under a complex Morse potential, where both mass and potential have imaginary components. In such non-Hermitian quantum systems, the reality or stability of eigenvalues and the probability density distribution can depend strongly on the interplay between $m_i$ and $a_i$.
	
	The Figure \ref{fig3a} is a colour plot with following Colour mapping:
	
	\textit{Blue region}: Allowed region where PPD is real and positive.
	
	\textit{Black region}: Forbidden or non-physical region (eigenfunction non-normalizable or unstable).
	
	\textit{Gray region}: Transitional region with negative value of PPD.
	
	It is observed that for positive $m_i$ and positive $a_i$ is depicted by the blue region indicating stability and real solutions. For negative $a_i$ and negative $m_i$, gray region is signified that points towards unstable states. Further at the origin, we observe a sharp transition line that alludes to a critical point separating stable and unstable regions. Hence, large positive imaginary mass $m_i$ and potential range $a_i$ contribute to stabilizing the probability density while the opposite signs lead to instabilities or non-normalizable eigenfunctions.
	
	The Figure \ref{fig3b} is plotted with extended range to express the behaviour of PPD with respect to $a_i$ and $m_i$. Sharp boundary near origin is observed along with tiny variations around ($m_i = 0$) or ($a_i = 0$) causing abrupt transitions. As in Figure \ref{fig3a}, stability of PPD is again found for positive-positive quadrant (blue) while the Instabilities persist for mixed-sign regions. It is conspicuous that the black regions are wider for small values that suggests nonlinearity or sensitivity of PPD near zero. This suggests that even small imaginary components can drastically alter the eigenfunction stability. Further, it also indicates non-Hermitian sensitivity of the PPD in the sense that a small parameter shift implies a large qualitative change in solution behaviour.
	
	In the third quadrant of the parameter space, where both $a_i$ and $m_i$ take negative values, the system exhibits a distinct and non-trivial behaviour. As indicated by the blue region in the general PPD plot (Figure \ref{fig3b}), a finite domain exists in which the probability density remains positive despite both parameters being negative. Notably, this physically admissible region appears only when the magnitude of the negative parameters becomes sufficiently large, implying that strong negative values of $a_i$ and $m_i$ are required to sustain a positive PPD. 
	
	The Imaginary Mass $m_i$ therefore, is a measure of a gain or loss term in the system. A Positive value of $m_i$ may represent gain (amplification), while negative $m_i$ indicates loss. Likewise, the imaginary Potential Range $a_i$ controls the extent of complex potential interaction, with a positive $a_i$ results in stabilizing effect thus enhancing localization. On the other hand, the negative $a_i$ results in destabilizing of PPD, leading to oscillatory or divergent solutions. Critical Boundaries can be identified in the extended complex phase space where the axes (($m_i=m_c$), ($a_i=a_c$)) are bifurcation lines separating stable and unstable domains. They might correspond to exceptional points, where eigenvalues coalesce or shift from real to complex.
	 
	The above-mentioned behaviour has far-reaching implication for Molecular Systems. For the Hydrogen molecule, stable eigenfunction behaviour occurs when both the mass modulation and potential modulation share the same sign and are sufficiently strong. On the other hand, mixed signs lead to non-normalizable states, signifying unphysical conditions to admit bound molecular states.
	
	Thus, the admissible parameter space for quantum behaviour is given by:
	
	$m_i \in (m_c, \infty)$, $a_i \in [0,a_c)$. Approaching these critical values, the area of existence collapses, and PPD diverges, signifying a transition from physically viable to physically unviable complex system.
	
	\subsection{Analysis of Real part of Energy Spectrum} \label{subsec5.4}
	
	Figure \ref{fig4} presents four distinct plots illustrating the variation of the real part of the energy $E_r$ for the hydrogen molecule as a function of the imaginary components of the reduced mass $m_i$ and the potential range $a_i$. These plots collectively examine the behaviour of the system governed by a complex mass under the influence of a complex Morse potential.
	
	The first plot (Figure \ref{fig4a}) is a color-coded parametric map depicting the dependence of $E_r$ on $m_i$ and $a_i$. The red region corresponds to the domain where real eigenvalues exist, representing physically admissible solutions, while the black region denotes the non-physical domain where Probability density become complex or negative. The white dashed lines at $m_i = 0$ and $a_i=0$ divide the parameter space into four quadrants. It is evident that real eigenvalues are confined to the region where both $m_i$ and $a_i$ are positive. As $a_i$ increases the permissible range of $m_i$ expands slightly, whereas no real solutions exist for negative values of either parameter. This plot reveals that the stability and reality of the eigenvalues are highly sensitive to the imaginary components of the system parameters. Consequently, real bound states are sustained only when both $m_i$ and $a_i$ assume positive values, emphasizing that the system’s physical validity is strongly constrained within a narrow region of the complex parameter space. The second plot, in Figure \ref{fig4b} is drawn for a wider range of imaginary parameters $a_i$ and $m_i$. It is observed that the gray region appear in the third quadrant which signifies the domain for the complex matter in which the real part of the eigenvalue is negative.
	
	The third plot (Figure \ref{fig4c}) illustrates the variation of $E_r$ with $m_i$ for several fixed values of $a_i$. The energy decreases monotonically with increasing $m_i$, exhibiting a sharp decline in the real part of eigenvalue, followed by saturation, where $E_r$ approaches a nearly constant value. A comparative analysis across different $a_i$ values shows that larger $a_i$ corresponds to higher energy levels for the same $m_i$. This behaviour indicates that while an increase in the imaginary component of the mass lowers the real part of the energy, the influence becomes negligible beyond a certain threshold. Moreover, an increase in $a_i$ leads to a consistent rise in $E_r$, suggesting that the imaginary component of the potential range contributes positively to the bound energy.
	
	The fourth plot in Figure \ref{fig4d} depicts the dependence of $E_r$ on $a_i$ for different fixed values of $m_i$, taken as integer multiples of the real mass component $m_r$. The curves exhibit a rapid increase in $E_r$ with $a_i$, with the growth rate becoming steeper as $m_i$ increases. At small $a_i$, a nearly flat energy region is observed, which transitions sharply into a divergent trend as $a_i$ increases. The inset provides the zoomed out version of the same plot, to demonstrate the abrupt increase in the value of $E_r$ for a critical value of ai is attained.    Physically, this indicates that increasing the imaginary part of the potential range significantly enhances the system’s energy, making the states less tightly bound. Furthermore, larger values of $m_i$ amplify this effect, demonstrating a strong coupling between the imaginary components of mass and potential in determining the system’s energy landscape.
	
	\subsection{Imaginary part of Energy Spectrum Analysis} \label{subsec5.5}
	
	The focus of the plots, shown in Figure \ref{fig5}, is to analyze the behaviour of the imaginary energy component $E_i$ with respect to the imaginary mass component $m_i$ and imaginary potential range $a_i$.
	
	Figure \ref{fig5a} is again a colour plot with the colour mapping defined as:
	
	\textit{Orange region}: Domain with real and positive valued $E_i$, implying physically meaningful energy spectrum.
	
	\textit{Gray region}: Reigion with negative values of $E_i$ implying growth in the eigenfunction of the complex system. 
	
	\textit{Black region}: Non-physical region - eigenfunction is non-normalizable and negative probability density. 
	
	Stability is observed (orange region) only in the first quadrant ($m_i > 0$, $a_i > 0$). As $a_i$ increases for a given $m_i > 0$, the orange region widens, indicating enhanced stability. For $m_i < 0$ or $a_i < 0$, the system remains unstable (black region). To interpret it physically, the imaginary component of the energy, $E_i$, is real and bounded only when both $m_i$ and $a_i$ are positive. Also, positive $m_i$ (gain-like mass component) and positive $a_i$ (localized complex potential) create a balanced non-Hermitian regime yielding quasi-bound states. On the other hand, negative or mixed signs lead to instability, representing non-normalizable unphysical eigenfunctions. The gray band is hemmed between orange and black region signifying decaying and unphysical states respectively. When the same plot is zoomed in Figure \ref{fig5b} for larger parametric range of $mi$ and $ai$, a gray region again appears in the third quadrant. 
	
	Figure \ref{fig5c} represents the variation of $E_i$ with $m_i$ (for fixed $a_i$) with the inset zooms into small $m_i$ region. It is noticed that $E_i$ decreases rapidly with increasing $m_i$ and then asymptotically approaches a small constant value. For larger $a_i$, the energy magnitude is higher at low $m_i$. It is further noted that the rate of decay in $E_i$ is sharper for larger $a_i$.
	
	The above behaviour can imply that the Increasing value of imaginary mass component $m_i$ suppresses the imaginary part of the energy. This suggests enhanced stability for larger imaginary mass, as $E_i$ decreases, meaning the system behaves more Hermitian-like. At small $m_i$, the non-Hermiticity is stronger larger $E_i$, leading to more significant gain/loss dynamics. Thus $a_i$ acts as a control parameter: higher values of $a_i$ implies stronger initial instability, but faster stabilization with increasing $m_i$.
	
	Figure \ref{fig5d} investigates the variation of $E_i$ with $a_i$ for fixed $m_i$ with the inset highlighting the small $a_i$ region. This imples that $E_i$ increases sharply beyond a critical $a_i$ for each $m_i$ with smaller $m_i$ values exhibiting lower thresholds and faster divergence. Alternatively, larger $m_i$ shifts the critical point to higher $a_i$, indicating enhanced stability.
	
	The points to the pivotal role of $a_i$ in deciding the non-Hermitian interaction range. As $a_i$ grows, it amplifies $E_i$, leading to non-physical growth. There exists a critical value of ai equalling ac for each $m_i$ beyond which $E_i$ diverges, thus representing a breakdown of quasi-bound state. Also, increasing value of $m_i$ raises the critical value $a_c$ leading to expansion of the stable domain, confirming stabilizing role of $m_i$.
	
	The emergence of large magnitudes in $E_i$ near critical values further corroborates the instability of the system in those regimes. The above results can be summarized in a tabular form as follows:
		
		\begin{table}[h!]
			\centering
			
			\renewcommand{\arraystretch}{1.15}
			
			\begin{tabular}{|>{\centering\arraybackslash}m{4cm}|
					>{\centering\arraybackslash}m{2cm}|
					>{\centering\arraybackslash}m{2cm}|
					>{\centering\arraybackslash}m{2cm}|
					>{\centering\arraybackslash}m{2cm}|}
				\hline
				\textbf{Behaviour of probability density $E_r$ and $E_i$}
				& \multicolumn{2}{c|}{\textbf{Fixed value of $a_i$}}
				& \multicolumn{2}{c|}{\textbf{Fixed value of $m_i$}} \\ \cline{2-5}
				
				\rule{0pt}{1.5em}
				& $m_i \rightarrow m_c$
				& $(m_c, \infty)$
				& $a_i \rightarrow a_c$
				& $(0,a_c)$ \\ \hline
				
				$E_r$ & Extremely high & Gradually decreases & Extremely high & Gradually increases \\ \hline
				$E_i$ & Extremely high & Gradually decreases & Extremely high & Gradually increases \\ \hline
				Area of probability density & Lowest & Increases & Lowest & Decreases \\ \hline
				Peak probability density & Extremely high & Gradually decreases & Extremely high & Gradually increases \\ \hline
			\end{tabular}
			
			\caption{Behaviour of ground state probablity density, energy eigen function with varying imaginary parameter of complex hydrogen ($H_2$) molecule in extended complex phase space. Here, $m_c$ and $a_c$ are critial values of imaginary mass and Morse parameter, respectively.}
		\end{table}

		\section{Reality of the spectrum} \label{sec6}
		
		\begin{figure}[h!]
			\centering
			
			\begin{subfigure}[b]{0.45\textwidth}
				\centering
				\includegraphics[width=\textwidth]{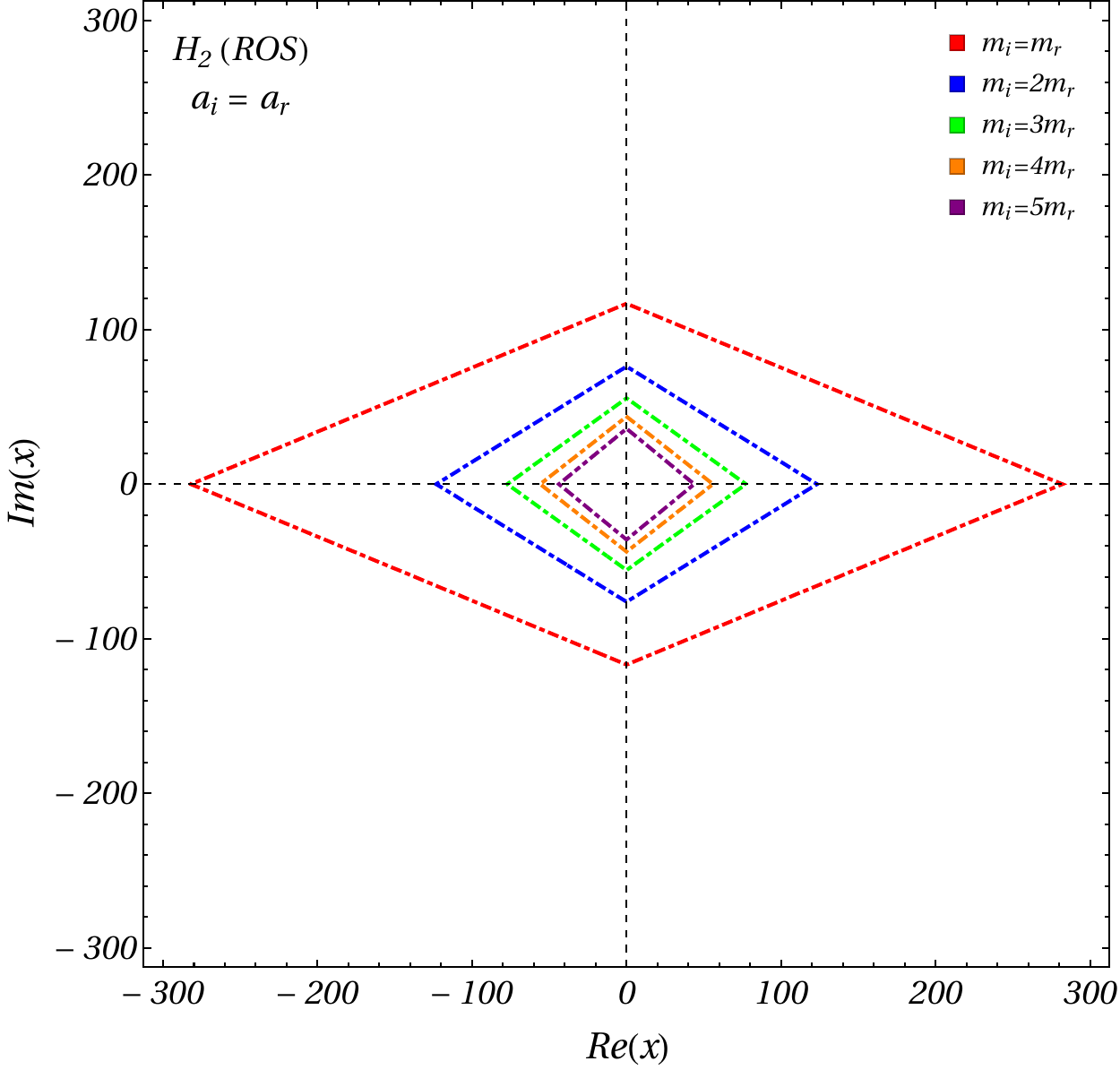}
				\caption{}
				\label{fig6a}
			\end{subfigure}
			\hfill
			\begin{subfigure}[b]{0.45\textwidth}
				\centering
				\includegraphics[width=\textwidth]{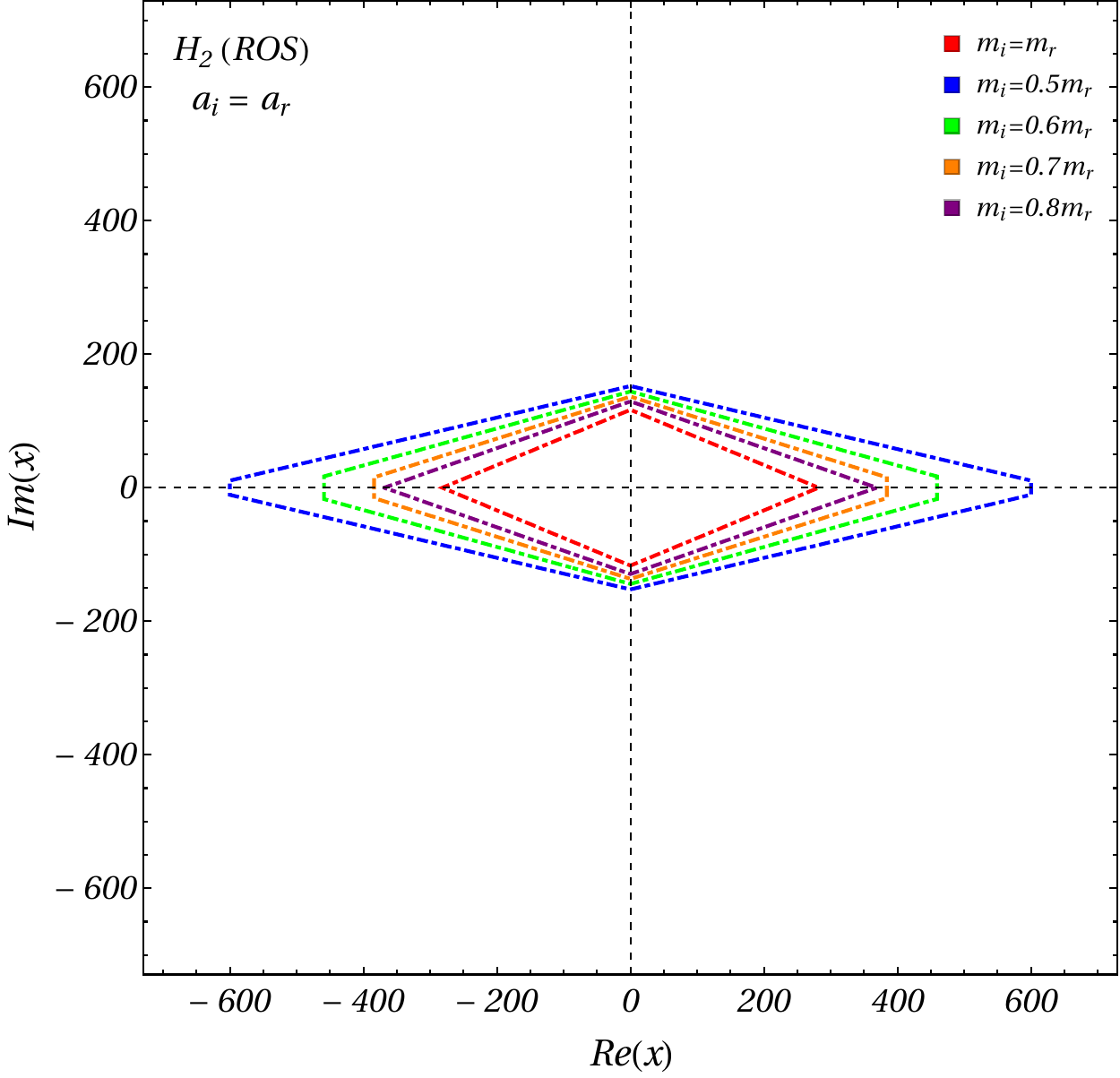}
				\caption{}
				\label{fig6b}
			\end{subfigure}
			
			\vspace{0.5cm}
			
			\begin{subfigure}[b]{0.45\textwidth}
				\centering
				\includegraphics[width=\textwidth]{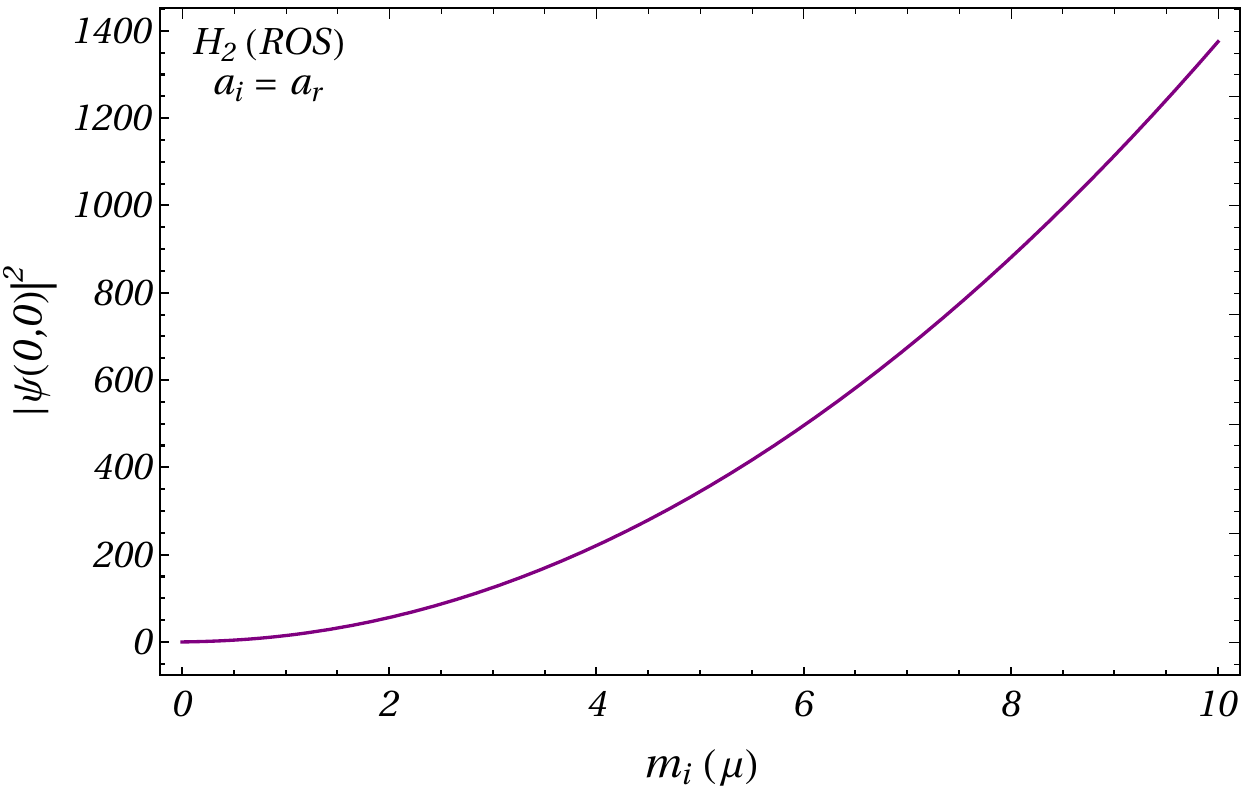}
				\caption{}
				\label{fig6c}
			\end{subfigure}
			\hfill
			\begin{subfigure}[b]{0.45\textwidth}
				\centering
				\includegraphics[width=\textwidth]{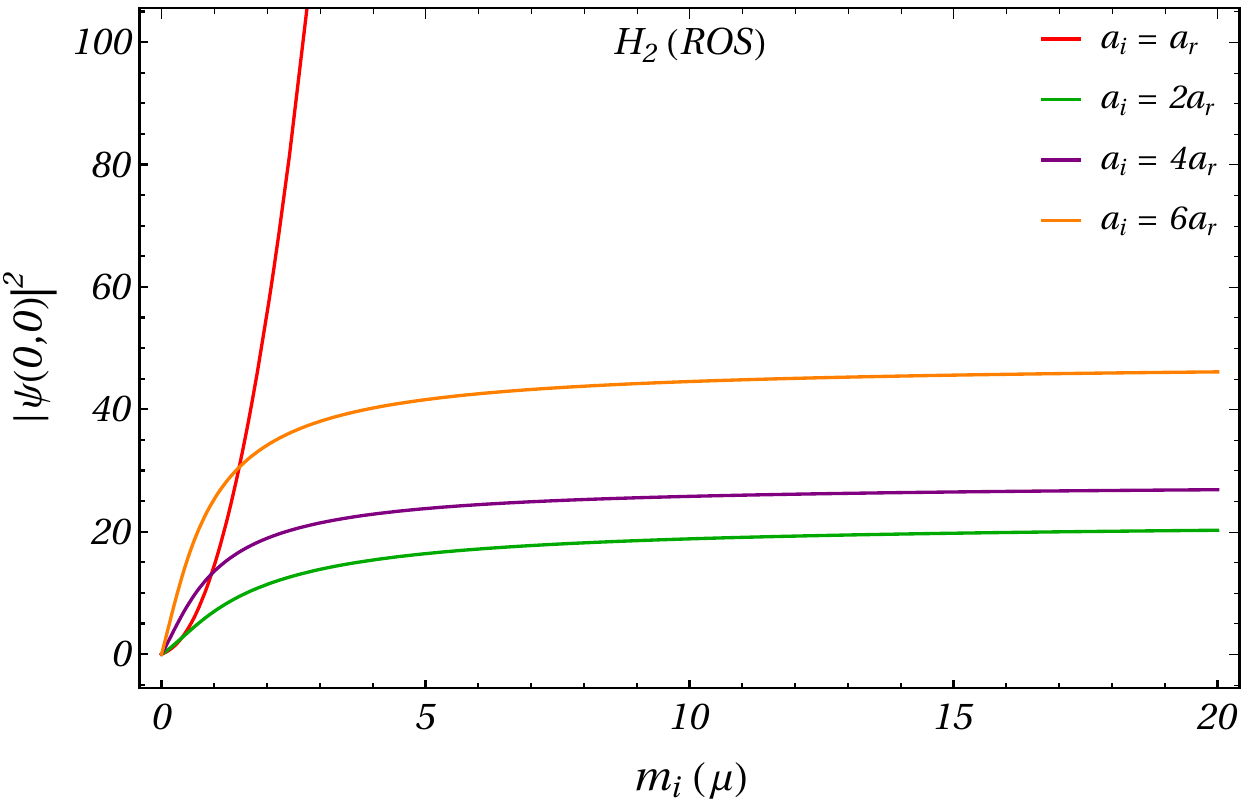}
				\caption{}
				\label{fig6d}
			\end{subfigure}
			
			\caption{The ground state probability density plots of the hydrogen ($H_2$) molecule with respect to $m_i$ for reality of spectrum (ROS) case. (a) and (b) reveals the variation of spartial confingement of probability density with varying $m_i$ and for the fixed value of $a_i$ taken as $a_i = a_r$. (b) shows variation of peak value of the probability density with respect to $m_i$ for the fixed $a_i$ as $a_i = a_r$. (d) shows dependency of peak probability density on complex mass parameter for various values of $a_i$. The value of real parameters for the $H_2$ molecule are taken as $V_{or} = 38266$ $cm^{-1}$, $a_r = 1.868$ $\AA^{-1}$, $m_r = 0.5039$ $\mu$.}
			\label{fig6}
		\end{figure}
		
		\begin{figure}[h!]
			\centering
			
			\begin{subfigure}[b]{0.45\textwidth}
				\centering
				\includegraphics[width=\textwidth]{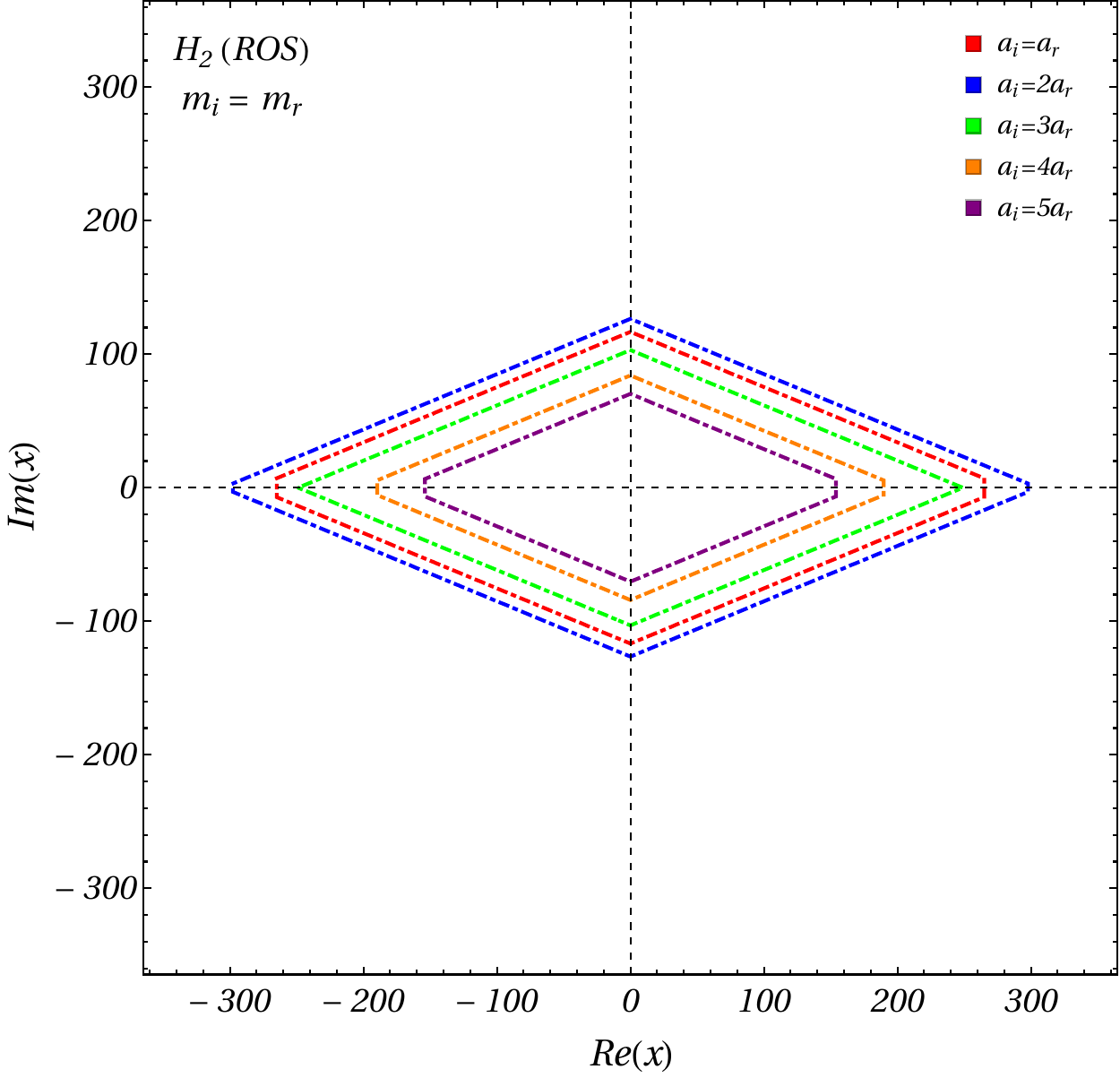}
				\caption{}
				\label{fig7a}
			\end{subfigure}
			\hfill
			\begin{subfigure}[b]{0.45\textwidth}
				\centering
				\includegraphics[width=\textwidth]{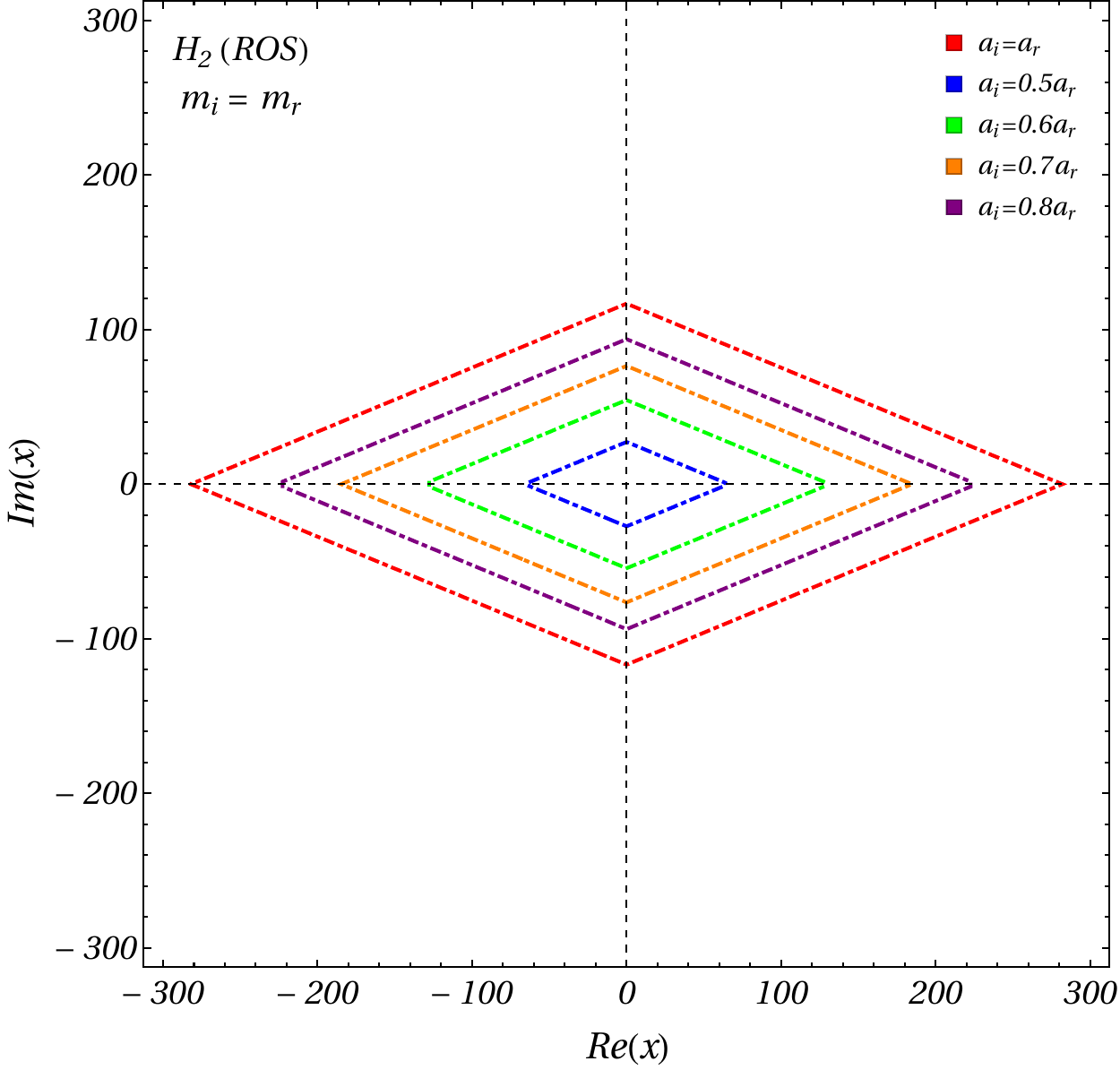}
				\caption{}
				\label{fig7b}
			\end{subfigure}
			
			\vspace{0.5cm}
			
			\begin{subfigure}[b]{0.45\textwidth}
				\centering
				\includegraphics[width=\textwidth]{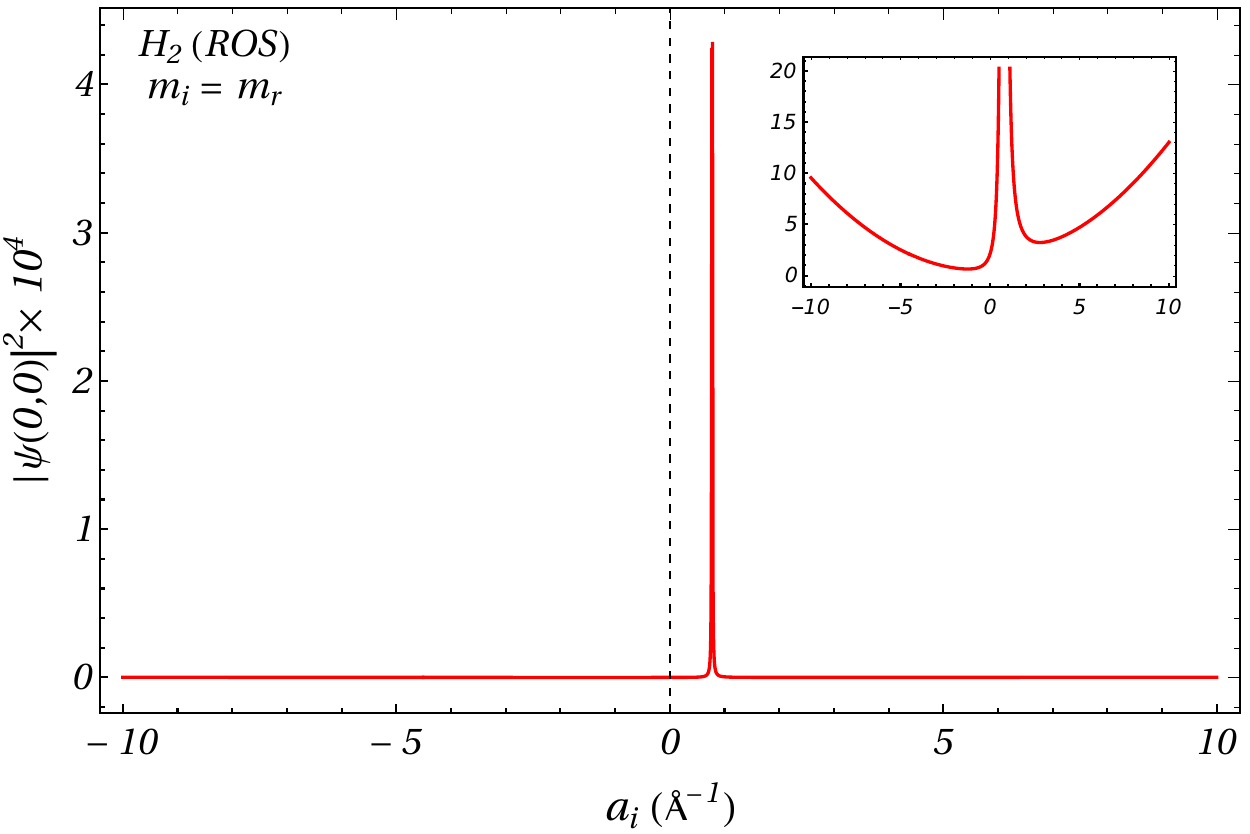}
				\caption{}
				\label{fig7c}
			\end{subfigure}
			\hfill
			\begin{subfigure}[b]{0.45\textwidth}
				\centering
				\includegraphics[width=\textwidth]{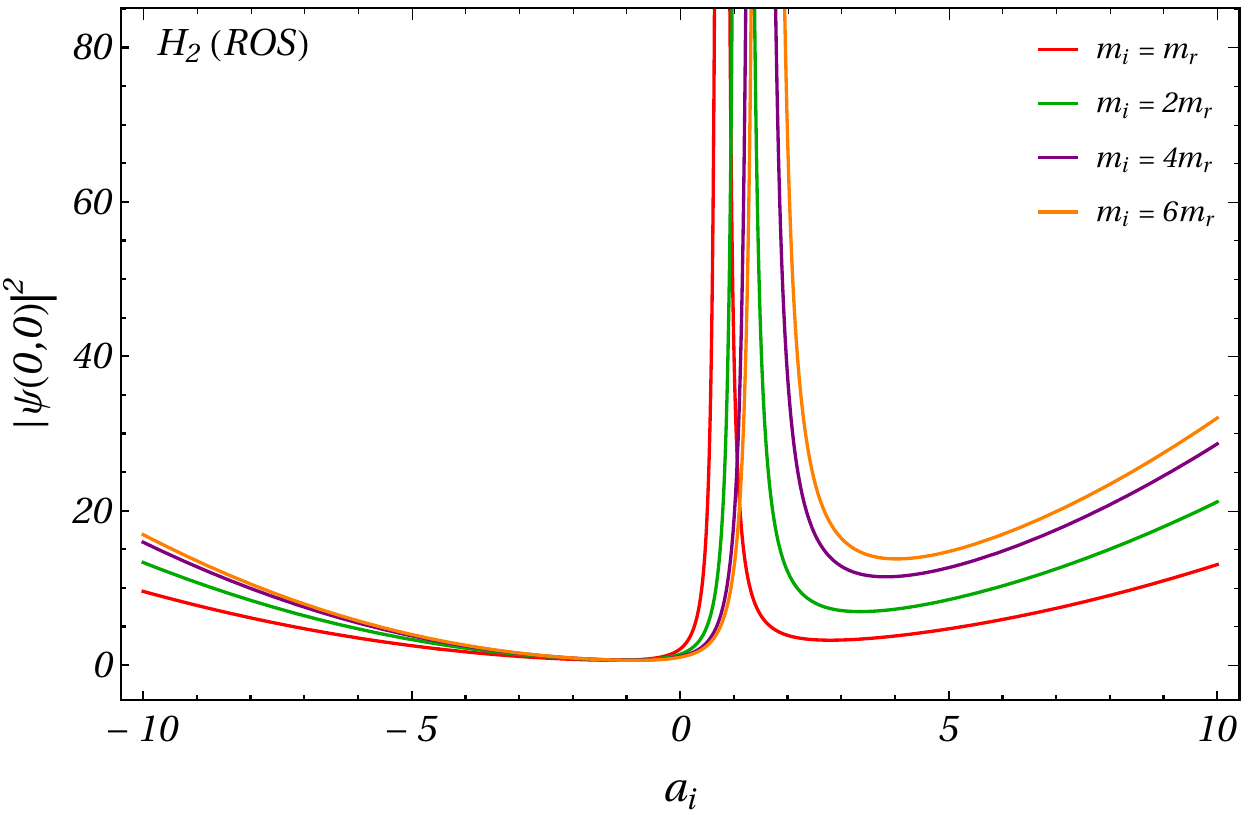}
				\caption{}
				\label{fig7d}
			\end{subfigure}
			
			\caption{The ground state probability density plots of the hydrogen ($H_2$) molecule with respect to $a_i$ for reality of spectrum (ROS) case. (a) and (b) reveals the variation of spartial confingement of probability density with varying $a_i$ and fixed value of $m_i$ taken as $m_i = m_r$.(c) shows variation of peak value of the probability density with respect to $a_i$ and fixed value of $m_i$, as $m_i = m_r$. (d) shows dependency of peak probability density on complex Morse parameter for various values of $m_i$. The value of real parameters for the $H_2$ molecule are taken as $V_{or} = 38266$ $cm^{-1}$, $a_r = 1.868$ $\AA^{-1}$, $m_r = 0.5039$ $\mu$.}
			\label{fig7}
		\end{figure}
		
		\begin{figure}[h!]
			\centering
			
			\begin{subfigure}[b]{0.45\textwidth}
				\centering
				\includegraphics[width=\textwidth]{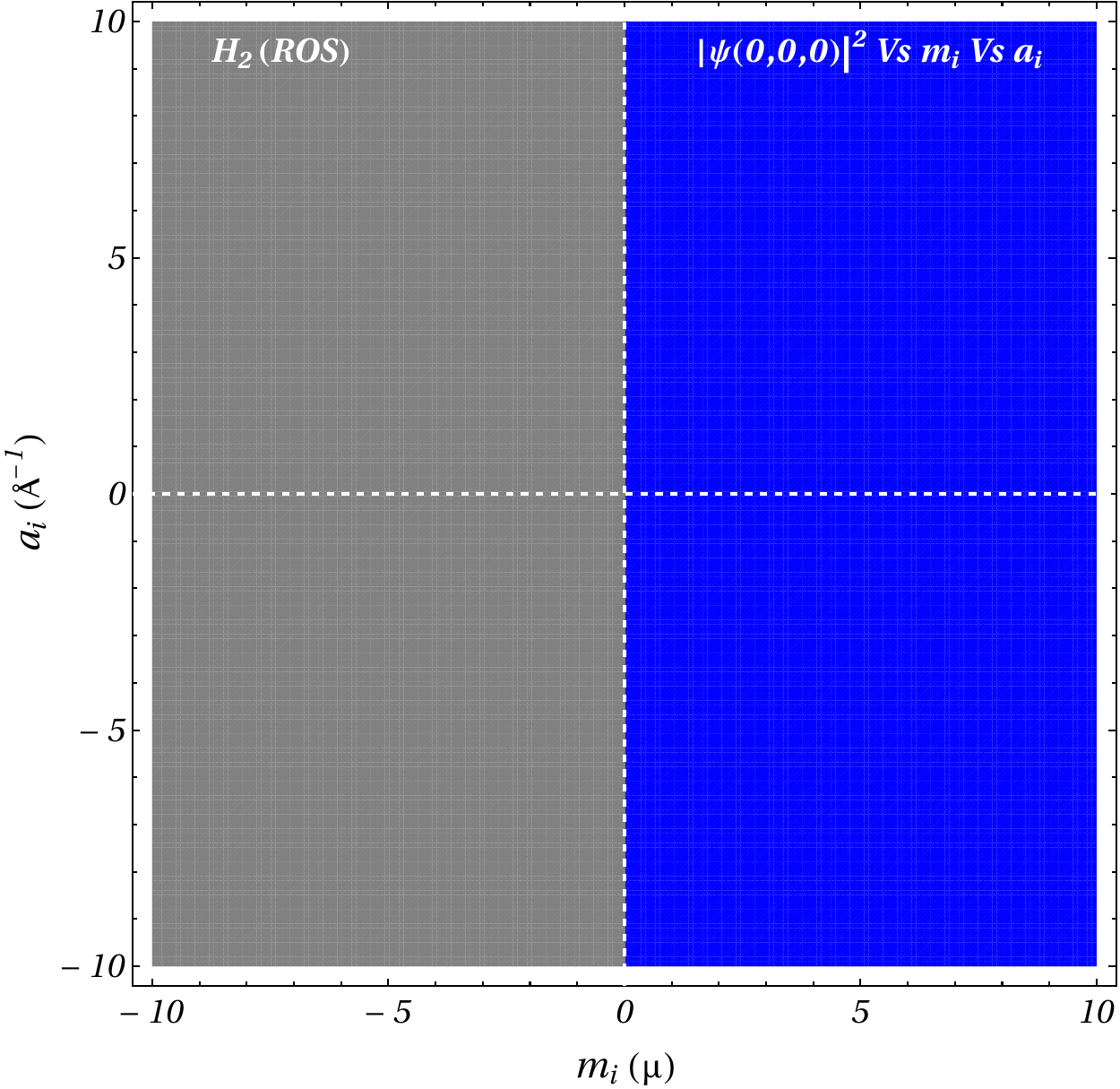}
				\caption{}
				\label{fig8a}
			\end{subfigure}
			\hfill
			\begin{subfigure}[b]{0.47\textwidth}
				\centering
				\includegraphics[width=\textwidth]{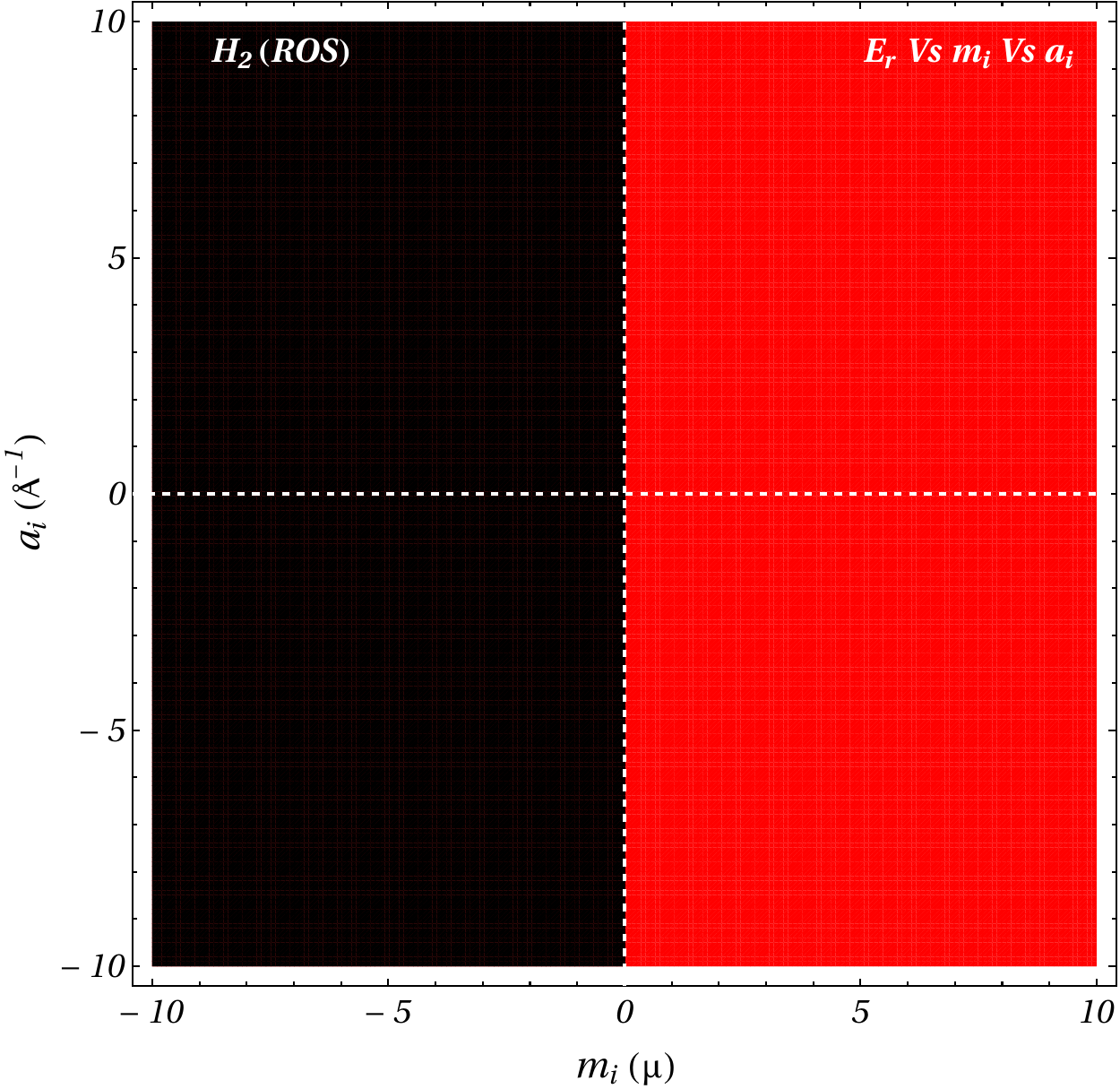}
				\caption{}
				\label{fig8b}
			\end{subfigure}
			
			\caption{The region plot of ground state peak probability density (PPD) and real energy eigenfunction, $E_r$ of the hydrogen ($H_2$) molecule in the parameteric space of ($a_i$, $m_i$) for reality of spectrum (ROS). (a) Blue shaded region shows the permissible values of ($a_i$, $m_i$) for which PPD is positive and normalizable whereas, gray region shows non-permissible region where PPD is negative. (b) The region shaded with red defines values of ($a_i$, $m_i$) for which $E_r$ is positive and black region corresponds to region with non-normalizable or negative PPD. Here, The value of real parameters for the $H_2$ molecule are taken as $V_{or} = 38266$ $cm^{-1}$, $a_r = 1.868$ $\AA^{-1}$, $m_r = 0.5039$ $\mu$.}
			\label{fig8}
		\end{figure}
		
		\begin{figure}[h!]
			\centering
			
			\begin{subfigure}[b]{0.3\textwidth}
				\centering
				\includegraphics[width=\textwidth]{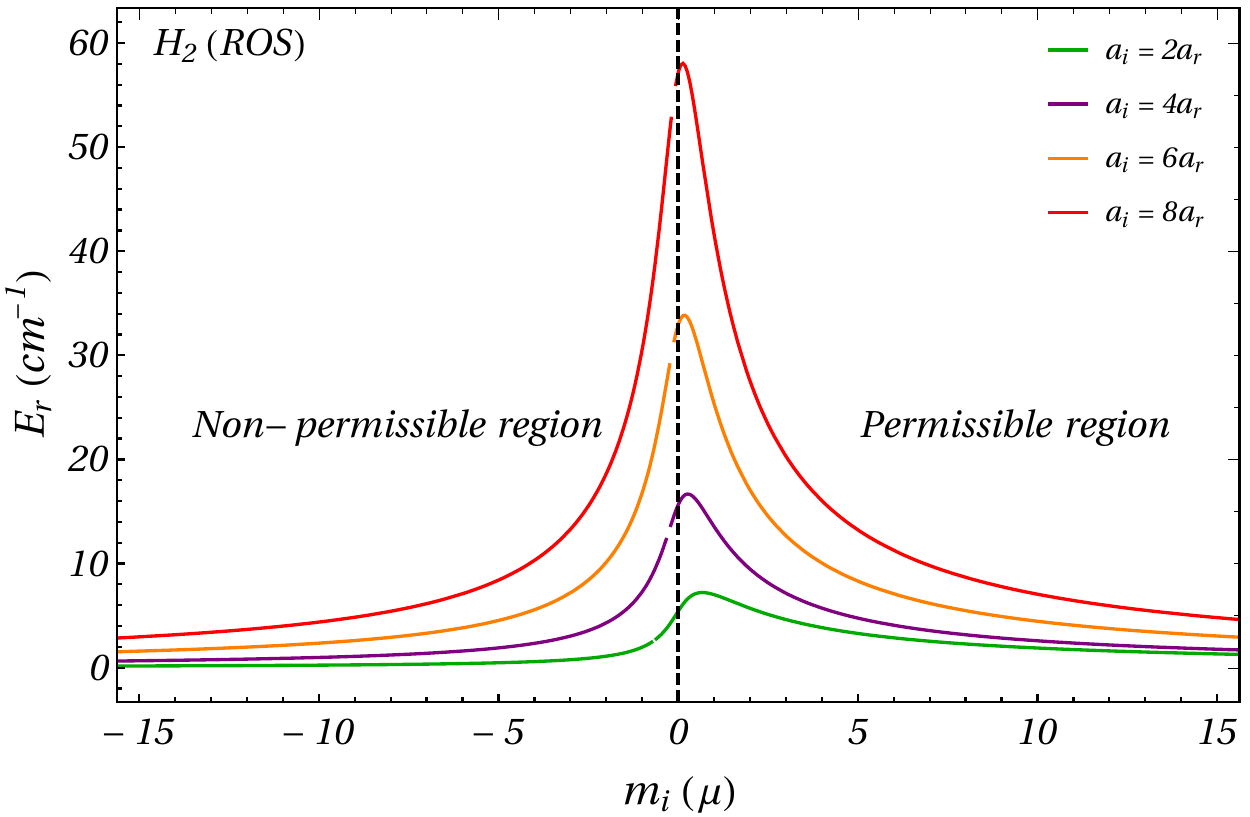}
				\caption{}
				\label{fig9a}
			\end{subfigure}
			\hfill
			\begin{subfigure}[b]{0.3\textwidth}
				\centering
				\includegraphics[width=\textwidth]{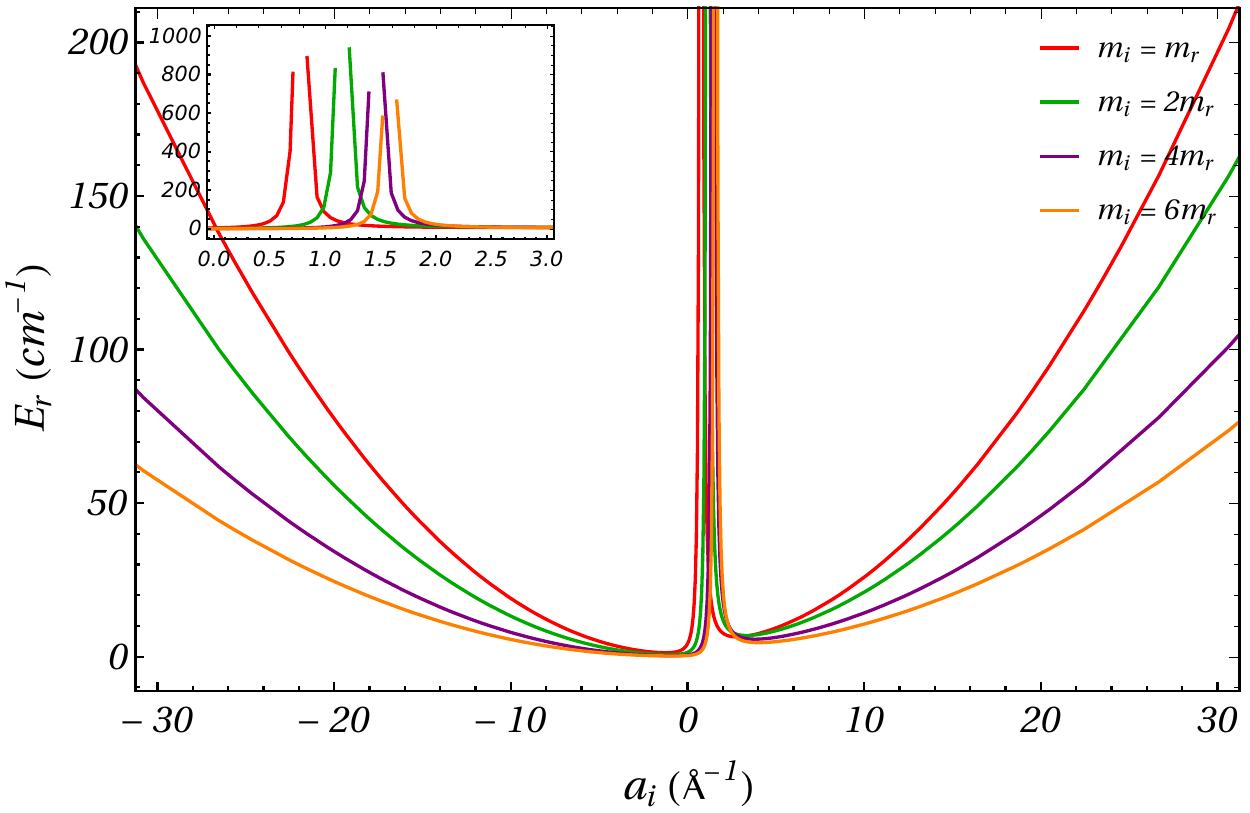}
				\caption{}
				\label{fig9b}
			\end{subfigure}
			\hfill
			\begin{subfigure}[b]{0.3\textwidth}
				\centering
				\includegraphics[width=\textwidth]{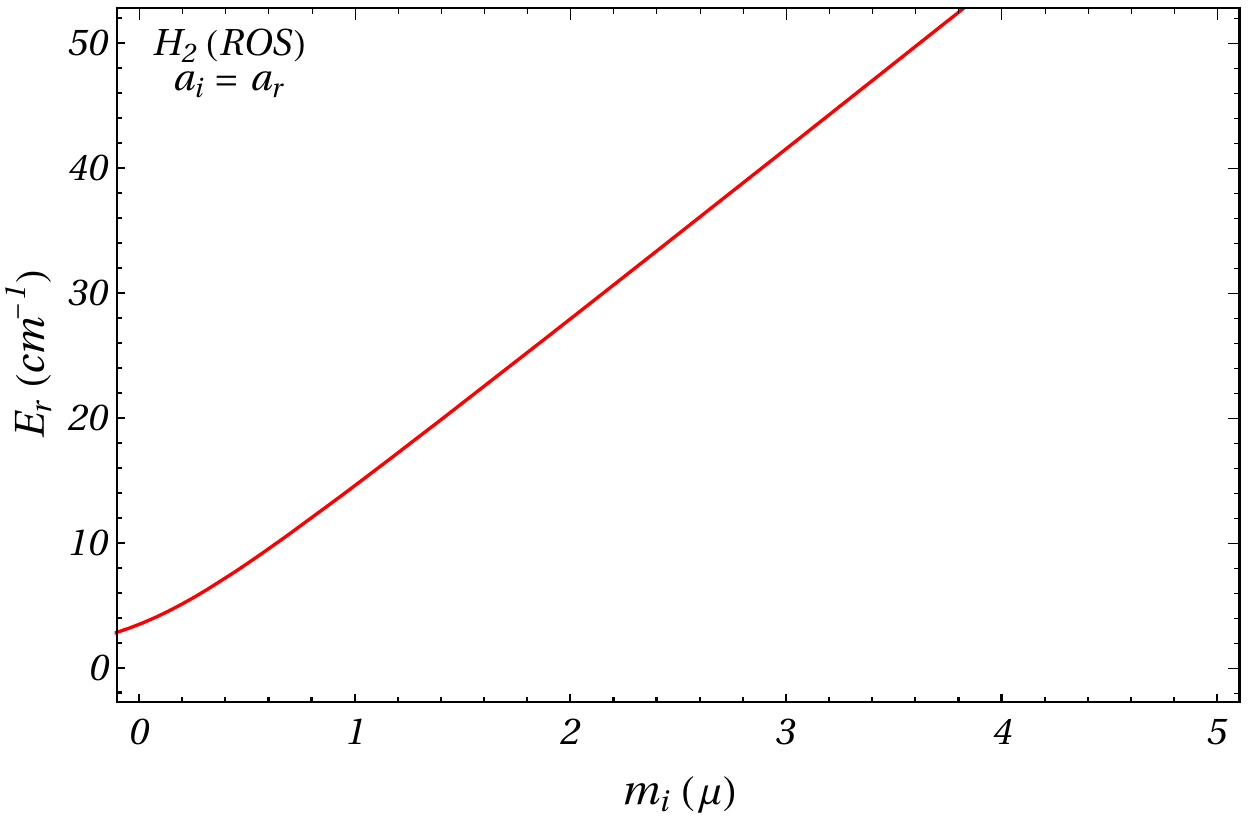}
				\caption{}
				\label{fig9c}
			\end{subfigure}
			
			\caption{The variation of ground state real energy eigenfunction with respect to imaginary parameters, $m_i$ and $a_i$, of the hydrogen ($H_2$) molecule for reality of spectrum (ROS). (a) shows the nature of $E_r$ with respect to $m_i$ for various values of $a_i$. Here, $m_i \ge 0$ and $m_i < 0$ defines permissible and non-permissible region, respectively. (b) reveals the nature of $E_r$ with respect to $a_i$ for various values of $m_i$. (a) shows the nature of $E_r$ with respect to $m_i$ for the fixed value of $a_i$, taken as $a_i = a_r$. The value of real parameters for the $H_2$ molecule are taken as $V_{or} = 38266$ $cm^{-1}$, $a_r = 1.868$ $\AA^{-1}$, $m_r = 0.5039$ $\mu$.}
			\label{fig9}
		\end{figure}
		
		In quantum mechanics, the reality of eigenspectra holds fundamental importance as it ensures the physical observability and stability of measurable quantities such as energy, momentum, and angular momentum. Real eigenvalues correspond to bound and physically meaningful states, which form the cornerstone of quantum theory and its predictive power. Recent advances in non-Hermitian quantum mechanics have revealed that even non-Hermitian systems can exhibit real and stable eigenspectra under specific symmetry conditions or a constraining relation among the potential parameters. This extension has profound implications, allowing the modelling of open, dissipative, or gain–loss systems without sacrificing physical realism. Understanding the conditions under which eigenvalues remain real is thus crucial for exploring novel quantum regimes, ensuring unitarity, and maintaining consistency between theoretical predictions and experimental observations.
		
		The added advantage of our formalism of solving the Schrödinger equation in extended complex phase space is that the imaginary value of the eigenspectra can be explicitly calculated in form of equation \ref{eq18b}. Putting the value of $E_i$ as zero in the said equation, two values of $\beta_3$ can be obtained as a constraining relation among the potential parameters as
		
		\begin{equation}
			A \beta_3^2 + B \beta_3 - \frac{A}{4} = 0, \label{eq28}
		\end{equation} 
		
		where $A = 2 \Omega m_r - (\Omega^2 - 1) m_i$ and $B = - 2 \Omega m_i - (\Omega^2 -1) m_r$; $\Omega = \frac{a_r}{a_i}$
		
		\begin{equation}
			\beta_3 = \frac{-B \pm (A^2 + B^2)^\frac{1}{2}}{2A}. \label{eq29}
		\end{equation}
		
		\subsection{Plots of the reality of spectrum (ROS) case} \label{subsec6.1}
		
		In complex matter, weak or indirect interactions and high delocalization are primarily governed by the imaginary components of the mass $m_i$ and the Morse potential $a_i$. In this section, we investigate the behaviour of the probability density and energy spectrum of the hydrogen molecule under the influence of complex Morse parameter $a_i$ and complex mass $m_i$. Particular emphasis is given to the cases involving imaginary components of the mass $m$ and Morse parameter $a$, as well as to the comparison between the general and “reality of spectrum” (ROS) cases. 
		
		\subsubsection{Probability Density variation with respect to $m_i$} \label{subsubsec6.11}
		
		Plot \ref{fig6} illustrates the probability density distribution of the hydrogen molecule as a function of the imaginary part of the mass, mi with $V_{or}$. A distinct transition in the region of existence is observed compared to the general case. As value of $m_i$ increases, the region of existence diminishes when the real and imaginary Morse parameters are equal. Inversely, the peak value of the probability density increases with increasing $m_i$.
		
		In Plot \ref{fig6d}, it is conspicuous that in the ROS case, for $a_i$ constant and $a_r$ not equal to $a_i$, the peak value of PPD remains almost unchanged with increasing value of $m_i$. This indicates that all the molecules with different masses have the same value of PPD indicating classical like deterministic physical system. Further,  a linear relationship is indicated between the probability density and the imaginary part of the mass when both real and imaginary part of the Morse parameters are equal.
		
		\subsubsection{Variation of Probability Density with Respect to $a_i$} \label{subsubsec6.12}
		
		Plot \ref{fig7} represents the variation of the probability density as a function of the imaginary Morse parameter, $a_i$ under the condition of equal real and imaginary mass. The analysis reveals a unique non-monotonic behaviour: the region of existence for the molecule initially expands with an increase in $a_i$, and subsequently contracts beyond a critical point. A similar pattern is observed for negative values of $a_i$. Additionally, the results indicate that the particle exhibits classical like behaviour at the lowest possible value of $a_i$, whereas higher imaginary components reintroduce quantum features.
		
		Figure \ref{fig7d} illustrates the dependence of the central probability density $|\psi(0,0)|^2$ of the $H_2$ system (ROS case) on the imaginary Morse parameter $a_i$ for a range of effective masses. Across all mass values, the probability density exhibits a pronounced divergence near $a_i=0$, indicating a strong resonance-like behaviour associated with the vanishing of the potential’s characteristic width. For negative $a_i$, the curves show a smooth monotonic rise, with heavier masses producing slightly higher probability densities suggesting an enhanced localization of the eigenfunction. On the positive side of $a_i$, the curves shift upward and broaden with increasing mass, reflecting stronger confinement arising from mass-dependent modifications of the effective potential. Overall, the figure demonstrates that both the sign and magnitude of $a_i$ critically influence the central density, while larger masses systematically amplify the sensitivity of $|\psi(0,0)|^2$ to variations in the Morse parameter.
		
		\subsubsection{Correlation between Peak Probability Density and Real Energy} \label{subsubsec6.13}
		
		Plot \ref{fig8} depicts the dependence of both the peak probability density and the real part of the energy on the parameters $m_i$ and $a_i$. The comparison between the ROS and the general cases reveals that the admissible range of $m_i$ and $a_i$ shifts significantly in the ROS condition. Notably, the imaginary mass cannot assume negative values, highlighting a physical constraint in maintaining the reality of the spectrum.
		
		\subsubsection{Real Energy Analysis} \label{subsubsec6.14}
		
		In Figure \ref{fig9a}, energy $E_r$ vs mass parameter $m_i$ shows strong, resonance-like structure cantered near $m_i=0$. As $m_i$ approaches zero the curves develop pronounced, asymmetric peaks whose height and sharpness grow with the Morse parameter $a_i$. Physically this means that when the imaginary part of the mass is small, the system is very sensitive to the potential shape: a larger $a_i$  concentrates the vibrational amplitude and pushes $E_r$ much higher in a narrow $m_i$ window, while smaller $a_i$ produces milder, broader responses. The inset emphasizes that for sufficiently large $|m_i|$ the energy falls off and that low-$a_i$ curves rise more gently, indicating that $m_i$ controls the inertial contribution to the eigenenergy.
		
		Figure \ref{fig9b} depicts energy $E_r$ vs Morse parameter $a_i$ complements this: all curves spike strongly near $a_i$ and then decay symmetrically for large $|a_i|$. Increasing $m_i$ shifts the entire family of curves downward and broadens the central feature, showing that a heavier imaginary mass suppresses high energy and smooths resonances induced by changes in the potential shape. Together $m_i$ and $a_i$ determine whether the system is in a delocalized, low-energy regime or a sharply resonant, highly localized regime which forms a key point for analyzing bound states and complex-mass effects in Morse potential.
		
		Further, the results reinforce the findings from the previous plots, emphasizing that the reality of the spectrum is maintained only in specific regions of the parameter space where the normalization condition holds. A distinct transition in the region of existence is observed when compared to the general case, indicating a clear dependence on the imaginary mass component.
		
		For the specific scenario (Figure \ref{fig9c}) where the real and imaginary Morse parameters are equal($a_r=a_i$), an increase in the imaginary mass leads to a reduction in the region of existence, signifying a contraction of the stable domain. Conversely, the peak value of the probability density exhibits an inverse trend, increasing progressively with higher imaginary mass. This behaviour suggests the presence of an approximately linear correlation between the probability density and the imaginary mass under conditions of equal real and imaginary Morse parameters. Moreover, the results imply that as the imaginary mass increases, the system progressively transitions toward classical-like behavior, indicating a diminishing influence of quantum fluctuations and an enhancement of deterministic dynamics.

		\section{Conclusions and Discussion} \label{sec7}
		
		In this study, we calculated the eigenvectors and corresponding eigenvalues for the complex Morse potential with a complex mass profile by solving the Schrödinger equation in complex phase space. The normalization condition for the eigenvectors was imposed through a two-dimensional integral in complex phase space,
		
		$$\int_{-\infty}^{\infty} \int_{-\infty}^{\infty} \psi^*\left(x_1,p_2\right) \psi\left(x_1,p_2\right) dx_1 dp_2 = \hbox{constant}.$$
		
		The normalized eigenfunctions were thereby obtained. In non-Hermitian systems, such a normalization is crucial as it ensures that expectation values of observables remain meaningful and that probabilities are well defined. We identified the conditions under which the eigenspectrum remains real and examined special cases of physical interest in detail. Notably, solving the Schrödinger equation in the complex phase space provides the advantage of directly extracting both the real and imaginary parts of the eigenspectrum in terms of explicit constraints on the potential parameters, an approach that is distinct from other techniques commonly used in the literature.
		
		It is evident from the study of eigenfunction normalization that the imaginary parts of the complex Morse potential parameters must be carefully chosen to ensure finite eigenvalues and definite positive eigenfunctions. Accordingly, the allowed values of the parameter $a_i$ are constrained to be positive and must satisfy the inequalities involving the $\beta_3$ variables defined in (\ref{eq25}). This constraint also permits the imaginary part of $V_{or}$ to be deduced directly from the relation given in Eq. (\ref{eq16}). 
		
		In this work, we have systematically analyzed the behaviour of the imaginary parts of the particle’s mass, the Morse parameter, and the corresponding energy eigenvalues for a quantum system characterized by complex mass under the action of a complex Morse potential. The investigation highlights the parameter domains where exact solutions exist, both for the general case and for the special case when the eigenvalue spectrum becomes real. A key finding is that, for exact solutions in the general case, the imaginary parts of the mass ($m_i$) and the Morse parameter ($a_i$) must simultaneously admit either positive or negative values. However, when the spectrum is restricted to real eigenvalues, $m_i$ is confined to positive values while $a_i$ can take both signs. Furthermore, the probability density $P$ remains real in the parameter domain defined by the blue regions of the plots, while it becomes complex in the black regions; this restriction is lifted in the real-spectrum case, where P remains always real.
		
		The interplay between $m_i$, $a_i$, and $P$ reveals nontrivial structures: in the general case, $P$ increases with either parameter when the other is held fixed, while for real spectra, $P$ becomes independent of $m_i$ when $a_i$ is fixed. Contour plots of $P$ further indicate that, although real and positive regions expand in the real case, the peak value of $P$ is significantly reduced.
		
		The relationship between the behaviour of imaginary parts of the mass of the particle and the energy eigenvalue in the general case show important feature. We take three cases based on the real and imaginary values of the Morse parameter as 
		
		Case I when $a_i > a_r$
		
		Case II when $a_i = a_r$
		
		Case III when $a_i < a_r$
		
		The relationship between $m_i$ and the energy eigenvalue exhibits distinct trends depending on the relative values of $a_i$ and its real part $a_r$ (Cases I–III). In all cases, as $m_i \rightarrow 0$, the imaginary part of the eigenvalue diverges in the general case, whereas the real part grows to a maximum before saturating. Conversely, in the real-spectrum case, the eigenvalue exhibits markedly different behaviors: it peaks and decreases in Cases I and III while increasing linearly in Case II. 
		
		In the general case, we analyzed the behaviour of the real and imaginary parts of the eigenvalue as functions of the imaginary component of the Morse parameter ($a_i$) It was observed that both the real and imaginary parts of the eigenvalue increase with $a_i$, indicating a net gain in the system. In contrast, when the spectrum is restricted to real values, the eigenspectrum assumes a parabolic form localized near the origin, which points to a coexistence of gain and loss mechanisms within the system.
		
		Further analysis of the probability density function $P$ with respect to the imaginary part of the mass ($m_i$) at the origin reveals distinct behaviour across the three cases considered. For Cases I and III in the general scenario, both $P$ and the imaginary part of the eigenvalue ($E_i$) diverge as $m_i \rightarrow 0$. In Case II, while $P$ remains finite, $E_i$ tends to infinity, implying that the imaginary part of the particle’s mass cannot vanish if the physical system is to exist. This trend, however, is reversed in the real-spectrum case: as $m_i \rightarrow 0$, both $P$ and $E_i$ remain finite and real. 
		
		\textit{Classification of Matter Based on Complex Parameters}
		
		These results thus establish that a complex-mass system under the influence of a complex Morse potential can generate five distinct categories of matter, each characterized by intrinsic physical properties. 
		
		It can thus be inferred from the plots \ref{fig1}-\ref{fig9} that increased value of $m_i$ spreads the particle’s kinetic profile, while moderate values of $a_i$ broadens the probability density within the potential well. The combined effect is a measure of spatial delocalization of the particle, and its interactions predominantly via long-range gravitational-like forces.
		
		The interaction of the hydrogen molecule with complex Morse potential parameters ($a = a_r + i a_i$) and complex mass ($m = m_r + i m_i$) reveals that different types of quantum matter can be postulated depending on the relative dominance and sign of their imaginary components and the corresponding reality of the energy spectrum. These classifications reflect distinct physical regimes that bridge real , quantum, quasi-classical, and non-physical domains.
		
		\textit{(a) Matter with Real eigenspectra}
		
		This form of matter corresponds to regions where both the normalization condition and reality of energy are satisfied, predominantly Quadrants I and IV of the Reality of Spectrum (ROS) Case. Here, $m_i$ admit only positive values, leading to finite, positive probability densities and purely real energy spectra. Physically, these states represent stable bound systems where particles behave as a Hermitian like matter, with well-defined spatial localization. The system’s dynamics are conservative, and there is no energy loss or decay. This matter corresponds to real, observable molecular states within the domain of non-Hermitian quantum mechanics.

		\textit{(b) Quasi-Stable Matter}		
		
		In the General Case, the boundary of grey and orange region depicted in Quadrants I in plot 5, exhibit a typical behaviour where both real and imaginary part of the energy eigenspectra $E_r$ and $E_i$ accepts finite values, with imaginary value $E_i$ taken as small and positive in the vicinity of zero. There exist regions where the normalization condition for the eigenfunction is satisfied with the probability density showing phase distortions or local amplitude modulations arising from the complex part of the potential. The probability envelope first increases with $m_i$ and then decreases, indicating fluctuations in the shape of the envelope. Though the energy remains almost real ($E_r > > E_i$), the probability density’s peak shifts, showing phase lag without total decoherence. Such states represent resonant or quasi-bound matter, where the system retains partial quantum coherence but exhibits energy dissipation due to non-zero $E_i$. These can be interpreted as metastable or decaying states, existing temporarily before transitioning into more stable configurations. The imaginary components $a_i$ and  $m_i$ effectively act as dissipation parameters, controlling the rate of decay and the localization width of the particle.
		
		\textit{(c) Purely Complex Matter}
		
		For regions such as Quadrant III in the General Case, where normalization is satisfied only for large negative values of the parameters, the energy eigenspectra admit negative real and imaginary parts. These states represent deeply quantum regimes, dominated by the effects of imaginary mass or potential parameters. Such matter behaves in a purely quantum mechanical sense, exhibiting non-Hermitian dynamics with oscillatory-decaying characteristics. Although not classically observable, these states can be relevant for describing quantum tunnelling, decaying molecular resonances, or open quantum systems interacting with an environment.
		
		\textit{(d) Non-Physical Matter}
		
		Regions where the normalization condition fails completely (Quadrants II and IV in the General Case and II and III in the ROS Case) correspond to non-physical or unobservable matter. In these regions, the probability density and both real and imaginary parts of the energy become undeterminable, leading to a breakdown of the quantum description. Such states can be viewed as mathematical artifacts arising from unbounded or non-normalizable eigenfunctions. The dominance of large ai and mi terms leads to non-conservative, divergent behaviour, making these states physically meaningless but useful for understanding the limits of spectral reality.
		
		This classification shows that imaginary mass and potential parameters not only modify the spectral properties but also dictate the nature of the resulting quantum matter ranging from real, observable molecular states to transient or non-physical configurations. The restoration of real spectra under ROS conditions thus marks the emergence of stable, interpretable matter, delineating a clear boundary between physical and non-physical regimes in complex quantum systems.
		
		\textit{(e) Matter with Determinate Probability Density exhibiting Classical Behaviour}
		
		Interestingly, the probability density of the hydrogen molecule becomes completely determinate and unchanging, indicating the loss of quantum interference and transition toward classical behaviour in regions where both the imaginary mass mi and the imaginary Morse parameter ai satisfy the following conditions:
		
		In the Imaginary mass Parameter Region, as shown in Plot 6(d), the probability density first expands and then becomes constant with increasing $m_i$. The unchanging or determinate regime occurs at the point of inflextion, where the increase in $m_i$ no longer affects the spatial probability profile. This typically happens when the real and imaginary part of the Morse papameters are not equal, where the eigenfunction behaves classically, showing no modulation or quantum oscillation.
		
		Therefore, the combined condition for a completely determinate and stationary probability density can thus be written as:
		
		$$E_i = 0, \quad a_i = \text{constant under the constraint $a_r \ne a_i$}$$
		
		In this parameter zone, the quantum wave character of the particle fades completely. The energy eigenvalues remain purely real, ensuring a constant probability density. The particle effectively behaves as a classical object, localized with constant probability in space, a signature of deterministic, non-quantum behaviour. In this “Reality of Spectrum” regime, both the normalization condition and real energy remain satisfied, the imaginary part of the energy vanishes, and the probability density becomes static, representing purely classical matter.
		
		We can interpret this region of determinate and unchanging probability density as a transitional or hybrid form of matter, which could conceptually represent a Quasi-Classical or “Meta-Stable” matter that exists in the limit where quantum oscillations freeze and the particle behaves as if trapped in a potential minimum. Further, the probability distribution becomes spatially static, implying the system no longer exhibits tunnelling or uncertainty-driven spreading. Such a matter form represents a quantum state that is deterministic, stable, and non-dissipative. Thus, theoretically, this region can correspond to a unique class of matter which is non-interacting, non-decaying, and deterministically stable, existing at the boundary between quantum and classical regimes.
		
		In summary, the nature of matter in systems governed by complex Morse-type potentials can be classified as follows:
		
		\begin{table}[h!]
			\centering
			\begin{tabular}{
					|>{\centering\arraybackslash}m{2.8cm}
					|>{\centering\arraybackslash}m{2.8cm}
					|>{\centering\arraybackslash}m{2.8cm}
					|>{\centering\arraybackslash}m{2.8cm}
					|>{\centering\arraybackslash}m{2.8cm}|}
				\hline
				\parbox[c][1.8cm][c]{2.8cm}{\centering \textbf{Type of Matter}} &
				\parbox[c][1.8cm][c]{2.8cm}{\centering \textbf{Dominant Parameter}} &
				\parbox[c][1.8cm][c]{2.8cm}{\centering \textbf{Energy Nature}} &
				\parbox[c][1.8cm][c]{2.8cm}{\centering \textbf{Normalization}} &
				\parbox[c][1.8cm][c]{2.8cm}{\centering \textbf{Physical Interpretation}} \\ \hline
				
				\parbox[c][1.8cm][c]{2.5cm}{\centering Real eigenspectral matter} &
				\parbox[c][1.8cm][c]{2.5cm}{\centering Positive $a_i$ and $m_i$} &
				\parbox[c][1.8cm][c]{2.5cm}{\centering Purely real} &
				\parbox[c][1.8cm][c]{2.5cm}{\centering Fully satisfied} &
				\parbox[c][1.8cm][c]{2.5cm}{\centering Stable, physical bound state} \\ \hline
				
				\parbox[c][1.8cm][c]{2.5cm}{\centering Quasi-Stable} &
				\parbox[c][1.8cm][c]{2.5cm}{\centering Moderate values $a_i$ and $m_i$} &
				\parbox[c][1.8cm][c]{2.5cm}{\centering Complex with small and finite $E_i$} &
				\parbox[c][1.8cm][c]{2.5cm}{\centering Fully satisfied} &
				\parbox[c][1.8cm][c]{2.5cm}{\centering Metastable, decaying resonance} \\ \hline
				
				\parbox[c][1.8cm][c]{2.5cm}{\centering Purely Complex} &
				\parbox[c][1.8cm][c]{2.5cm}{\centering Large negative parameters} &
				\parbox[c][1.8cm][c]{2.5cm}{\centering Negative $E_r$ and $E_i$} &
				\parbox[c][1.8cm][c]{2.5cm}{\centering Satisfied in a narrow range} &
				\parbox[c][1.8cm][c]{2.5cm}{\centering Deeply quantum, tunelling states} \\ \hline
				
				\parbox[c][1.8cm][c]{2.5cm}{\centering Non-Physical} &
				\parbox[c][1.8cm][c]{2.5cm}{\centering Large $a_i$ and $m_i$} &
				\parbox[c][1.8cm][c]{2.5cm}{\centering Undeterminable} &
				\parbox[c][1.8cm][c]{2.5cm}{\centering Not satisfied} &
				\parbox[c][1.8cm][c]{2.5cm}{\centering Non-physical divergent states} \\ \hline
				
				\parbox[c][1.8cm][c]{2.5cm}{\centering Deterministic Matter} &
				\parbox[c][1.8cm][c]{2.5cm}{\centering $E_i = 0,$ $a_i = \text{constant}$, $a_r \ne a_i$} &
				\parbox[c][1.8cm][c]{2.5cm}{\centering Purely Real} &
				\parbox[c][1.8cm][c]{2.5cm}{\centering Fully Satisfied} &
				\parbox[c][1.8cm][c]{2.5cm}{\centering Static matter-like field} \\ \hline
			\end{tabular}
			\caption{Classification of matter and their discription for a complex quantum system under complex Morse potential in extended complex phase space on the basis of imaginary parameters, $m_i$ and $a_i$}
		\end{table}
		
		The classification of matter arising from the interplay of complex mass parameters ($m = m_r + i m_i$) and complex Morse potential parameters ($a = a_r + i a_i$) reveals a spectrum of physical regimes ranging from observable molecular states to exotic, non-Hermitian matter forms. The five identified classes of matter may be interpreted as distinct phases of a single quantum system governed by complex mass and Morse parameters. As the control parameters ($m_i, a_i$) vary, the system undergoes qualitative transitions in spectral reality, probability localization, and dynamical behaviour. These transitions delineate Hermitian-like, resonant, deeply quantum, non-physical, and quasi-classical phases, analogous to phase transitions in non-Hermitian quantum mechanics. Thus, the classification may be viewed as a phase diagram of matter emerging from a unified complex-parameter framework.
		
		Each class captures a distinct limit of quantum behaviour and suggests unique theoretical or phenomenological applications. These matter states can be physically interpreted as follows:
		
		(a) Matter with Real eigenspectra
		
		This class corresponds to fully stable, normalisable states with purely real energy eigenvalues and finite probability density. Such matter exhibits conservative dynamics, spatial localisation, and bounded motion. These characteristics make it analogous to real molecular systems, with potential application in modelling non-Hermitian corrections to bound states and effective molecular potentials
		
		(b) Quasi-Stable Matter
		
		States with small but finite imaginary energy components can represent metastable or resonant configurations. Their wavefunctions exhibit modulated envelopes and limited phase coherence, reflecting partial decay or energy leakage. This regime may be relevant for modelling open quantum systems, short-lived molecular resonances, quantum tunnelling processes, and even decaying dark-sector models where long lifetimes ($E_i \rightarrow 0^+$) mimic “slow-decay” cosmological components.
		
		(c) Purely Complex Matter
		
		When both real and imaginary parts of the energy become negative yet normalisability persists for extreme parameter values, the system enters a deeply non-Hermitian, oscillatory-decaying domain. Such states can behave like strongly interacting quantum modes, making them suitable for representing tunnelling-dominated regimes, quantum dissipation, induced decoherence, and non-Hermitian extensions of quantum transport. They may provide a framework for exploring phenomena where classical descriptions fail and quantum interference persists despite loss channels.
		
		(d) Non-Physical Matter.
		
		Regions where normalization breaks down represent non-physical or divergent states. Although not physically realizable, they may serve an important role in identifying spectral boundaries, instability thresholds, and conditions for loss of Hermiticity. These regions help delineate when physical observables cease to exist, thereby offering guidance for stability analysis, parameter-space constraints, and renormalization-like considerations in complex quantum theories.
		
		(e) Quasi-Classical or Determinate Matter.
		
		In parameter regimes admitted in this case, the probability density acquires a static, determinate form. Quantum oscillations freeze out, leading to spatially stable and effectively classical behaviour. This matter is non-dissipative, non-interacting, and long-lived. Its delocalised yet time-independent probability profile can suggest applications in semiclassical modelling, quantum–classical transition studies, coherence-controlled molecular states, and notably as a theoretical analogue of dark matter, owing to its collisionless, stable, and gravitational-like interaction characteristics.
		
		In summary, these five categories represent a continuum of behaviour governed by the balance of real and imaginary components of mass and potential parameters. From real, observable molecular states to quasi-classical, dark-matter–like candidates, the framework presents a unified approach to understanding how complex quantum parameters can generate qualitatively distinct forms of matter, each with its own theoretical significance and possible physical applications.
		
		\bibliographystyle{unsrt}
		\bibliography{References.bib}

\begin{thebibliography}{10}

\bibitem{bender2007making}
Carl~M Bender.
\newblock Making sense of non-hermitian hamiltonians.
\newblock {\em Reports on Progress in Physics}, 70(6):947--1018, 2007.

\bibitem{bender1999PT}
Carl~M Bender, Stefan Boettcher, and Peter~N Meisinger.
\newblock PT-symmetric quantum mechanics.
\newblock {\em Journal of Mathematical Physics}, 40(5):2201--2229, 1999.

\bibitem{bender2005introduction}
Carl~M Bender.
\newblock Introduction to PT-symmetric quantum theory.
\newblock {\em Contemporary physics}, 46(4):277--292, 2005.

\bibitem{bender2004extension}
Carl~M Bender, Dorje~C Brody, and Hugh~F Jones.
\newblock Extension of pt-symmetric quantum mechanics to quantum field theory
  with cubic interaction.
\newblock {\em Physical Review D}, 70(2):025001, 2004.

\bibitem{bender1998real}
Carl~M Bender and Stefan Boettcher.
\newblock Real spectra in non-hermitian hamiltonians having p t symmetry.
\newblock {\em Physical review letters}, 80(24):5243, 1998.

\bibitem{mostafazadeh2010pseudo}
Ali Mostafazadeh.
\newblock Pseudo-hermitian representation of quantum mechanics.
\newblock {\em International Journal of Geometric Methods in Modern Physics},
  7(07):1191--1306, 2010.

\bibitem{takata2017pt}
Kenta Takata and Masaya Notomi.
\newblock Pt-symmetric coupled-resonator waveguide based on buried
  heterostructure nanocavities.
\newblock {\em Physical Review Applied}, 7(5):054023, 2017.

\bibitem{ozdemir2019parity}
{\c{S}}ahin~Kaya {\"O}zdemir, Stefan Rotter, Franco Nori, and Lan Yang.
\newblock Parity--time symmetry and exceptional points in photonics.
\newblock {\em Nature materials}, 18(8):783--798, 2019.

\bibitem{longhi2017parity}
Stefano Longhi.
\newblock Parity-time symmetry meets photonics: A new twist in non-hermitian
  optics.
\newblock {\em Europhysics Letters}, 120(6):64001, 2017.

\bibitem{miri2019exceptional}
Mohammad-Ali Miri and Andrea Alu.
\newblock Exceptional points in optics and photonics.
\newblock {\em Science}, 363(6422):eaar7709, 2019.

\bibitem{ding2022non}
Kun Ding, Chen Fang, and Guancong Ma.
\newblock Non-hermitian topology and exceptional-point geometries.
\newblock {\em Nature Reviews Physics}, 4(12):745--760, 2022.

\bibitem{li2024observation}
Zhen Li, Li-Wei Wang, Xulong Wang, Zhi-Kang Lin, Guancong Ma, and Jian-Hua
  Jiang.
\newblock Observation of dynamic non-hermitian skin effects.
\newblock {\em Nature Communications}, 15(1):6544, 2024.

\bibitem{lin2022topological}
Quan Lin, Tianyu Li, Lei Xiao, Kunkun Wang, Wei Yi, and Peng Xue.
\newblock Topological phase transitions and mobility edges in non-hermitian
  quasicrystals.
\newblock {\em Physical Review Letters}, 129(11):113601, 2022.

\bibitem{hurst2022non}
Hilary~M Hurst and Benedetta Flebus.
\newblock Non-hermitian physics in magnetic systems.
\newblock {\em Journal of Applied Physics}, 132(22), 2022.

\bibitem{lin2011unidirectional}
Zin Lin, Hamidreza Ramezani, Toni Eichelkraut, Tsampikos Kottos, Hui Cao, and
  Demetrios~N Christodoulides.
\newblock Unidirectional invisibility induced by pt-symmetric periodic
  structures.
\newblock {\em Physical Review Letters}, 106(21):213901, 2011.

\bibitem{feng2014single}
Liang Feng, Zi~Jing Wong, Ren-Min Ma, Yuan Wang, and Xiang Zhang.
\newblock Single-mode laser by parity-time symmetry breaking.
\newblock {\em Science}, 346(6212):972--975, 2014.

\bibitem{lee2009observation}
Sang-Bum Lee, Juhee Yang, Songky Moon, Soo-Young Lee, Jeong-Bo Shim, Sang~Wook
  Kim, Jai-Hyung Lee, and Kyungwon An.
\newblock Observation of an exceptional point in a chaotic optical microcavity.
\newblock {\em Physical review letters}, 103(13):134101, 2009.

\bibitem{takata2021observing}
Kenta Takata, Kengo Nozaki, Eiichi Kuramochi, Shinji Matsuo, Koji Takeda,
  Takuro Fujii, Shota Kita, Akihiko Shinya, and Masaya Notomi.
\newblock Observing exceptional point degeneracy of radiation with electrically
  pumped photonic crystal coupled-nanocavity lasers.
\newblock {\em Optica}, 8(2):184--192, 2021.

\bibitem{chen2017exceptional}
Weijian Chen, {\c{S}}ahin Kaya~{\"O}zdemir, Guangming Zhao, Jan Wiersig, and
  Lan Yang.
\newblock Exceptional points enhance sensing in an optical microcavity.
\newblock {\em Nature}, 548(7666):192--196, 2017.

\bibitem{qin2026photonic}
Haoye Qin, Wenjing Lv, Zhe Zhang, Zijin Yang, Jue Li, Mengyao Li, Bo~Li,
  Ji~Zhou, Romain Fleury, Patrice Genevet, et~al.
\newblock Photonic exceptional points in engineered materials and their
  emerging applications.
\newblock {\em Nature Reviews Materials}, 11(6):453--468, 2026.

\bibitem{li2023coherent}
Zhifeng Li, Hai Lin, Rongxin Tang, Haitao Chen, Jiaru Tang, Rui Zhou, Jing Jin,
  and Yangjie Liu.
\newblock Coherent perfect absorber and laser induced by directional emissions
  in the non-hermitian photonic crystals.
\newblock {\em Optics Express}, 31(25):41276--41291, 2023.

\bibitem{zhu2014pt}
Baogang Zhu, Rong L{\"u}, and Shu Chen.
\newblock Pt symmetry in the non-hermitian su-schrieffer-heeger model with
  complex boundary potentials.
\newblock {\em Physical Review A}, 89(6):062102, 2014.

\bibitem{xing2017spontaneous}
Yan Xing, Lu~Qi, Ji~Cao, Dong-Yang Wang, Cheng-Hua Bai, Hong-Fu Wang, Ai-Dong
  Zhu, and Shou Zhang.
\newblock Spontaneous pt-symmetry breaking in non-hermitian coupled-cavity
  array.
\newblock {\em Physical Review A}, 96(4):043810, 2017.

\bibitem{jin2009solutions}
L~Jin and Z~Song.
\newblock Solutions of pt-symmetric tight-binding chain and its equivalent
  hermitian counterpart.
\newblock {\em Physical Review A—Atomic, Molecular, and Optical Physics},
  80(5):052107, 2009.

\bibitem{kaushal2002quantum}
RS~Kaushal and Parthasarathi.
\newblock Quantum mechanics of complex hamiltonian systems in one dimension.
\newblock {\em Journal of physics A: Mathematical and General},
  35(41):8743--8761, 2002.

\bibitem{bagchi2015bicomplex}
Bijan Bagchi and Abhijit Banerjee.
\newblock Bicomplex hamiltonian systems in quantum mechanics.
\newblock {\em Journal of Physics A: Mathematical and Theoretical},
  48(50):505201, 2015.

\bibitem{yang2005quantum}
Ciann-Dong Yang.
\newblock Quantum dynamics of hydrogen atom in complex space.
\newblock {\em Annals of Physics}, 319(2):399--443, 2005.

\bibitem{yang2007quantum}
Ciann-Dong Yang.
\newblock Quantum motion in complex space.
\newblock {\em Chaos, Solitons \& Fractals}, 33(4):1073--1092, 2007.

\bibitem{parthasarathi2004complex}
Parthasarathi, D~Parashar, and RS~Kaushal.
\newblock Complex phase space formulation of supersymmetric quantum mechanics:
  analysis of shape-invariant potentials.
\newblock {\em Journal of Physics A: Mathematical and General}, 37(3):781--801,
  2004.

\bibitem{bhardwaj2016quantum}
SB~Bhardwaj, Ram~Mehar Singh, and SC~Mishra.
\newblock Quantum mechanics of pt and non-pt-symmetric potentials in three
  dimensions.
\newblock {\em Pramana}, 87(1):10, 2016.

\bibitem{bhardwaj2014eigenspectra}
SB~Bhardwaj, Ram~Mehar Singh, and SC~Mishra.
\newblock Eigenspectra of a complex coupled harmonic potential in three
  dimensions.
\newblock {\em Computers \& Mathematics with Applications}, 68(12):2068--2079,
  2014.

\bibitem{parthasarathi2003quantum}
Parthasarathi and RS~Kaushal.
\newblock Quantum mechanics of complex sextic potentials in one dimension.
\newblock {\em Physica Scripta}, 68(2):115--127, 2003.

\bibitem{chand2009solution}
Fakir Chand, SC~Mishra, and Ram~Mehar Singh.
\newblock Solution of an analogous schr{\"o}dinger equation for-symmetric
  sextic potential in two dimensions.
\newblock {\em Pramana}, 73(2):349--361, 2009.

\bibitem{singh2009solution}
Ram~Mehar Singh, Fakir Chand, and SC~Mishra.
\newblock Solution of schr{\"o}dinger equation for two-dimensional complex
  quartic potentials.
\newblock {\em Communications in Theoretical Physics}, 51(3):397--406, 2009.

\bibitem{singh2014solving}
Ram~Mehar Singh.
\newblock On solving the schr{\"o}dinger equation for a complex deictic
  potential in one dimension.
\newblock {\em Pramana}, 83(3):301--316, 2014.

\bibitem{singh2013quantum}
Ram~Mehar Singh.
\newblock Quantum mechanics of complex octic potential in one dimension.
\newblock {\em Journal of Quantum Information Science}, 3(1):42, 2013.

\bibitem{ahmed2001real}
Zafar Ahmed.
\newblock Real and complex discrete eigenvalues in an exactly solvable
  one-dimensional complex pt-invariant potential.
\newblock {\em Physics Letters A}, 282(6):343--348, 2001.

\bibitem{bagchi2000sl}
Bijan Bagchi and Christiane Quesne.
\newblock sl (2, c) as a complex lie algebra and the associated non-hermitian
  hamiltonians with real eigenvalues.
\newblock {\em Physics Letters A}, 273(5-6):285--292, 2000.

\bibitem{znojil1999pt}
Miloslav Znojil.
\newblock Pt- symmetric harmonic oscillators.
\newblock {\em Physics Letters A}, 259(3-4):220--223, 1999.

\bibitem{dutt1988supersymmetry}
Ranabir Dutt, Avinash Khare, and Uday~P Sukhatme.
\newblock Supersymmetry, shape invariance, and exactly solvable potentials.
\newblock {\em American Journal of Physics}, 56(2):163--168, 1988.

\bibitem{weisskopf1930berechnung}
Victor Weisskopf and Eugene Wigner.
\newblock Berechnung der nat{\"u}rlichen linienbreite auf grund der diracschen
  lichttheorie.
\newblock {\em Zeitschrift f{\"u}r Physik}, 63(1):54--73, 1930.

\bibitem{mostafazadeh2020time}
Ali Mostafazadeh.
\newblock Time-dependent pseudo-hermitian hamiltonians and a hidden geometric
  aspect of quantum mechanics.
\newblock {\em Entropy}, 22(4):471, 2020.

\bibitem{guo2024non}
Wu-zhong Guo and Tao Liu.
\newblock Non-hermitian spacetime and generalized thermofield double formalism.
\newblock {\em arXiv preprint arXiv:2406.06961}, 2024.

\bibitem{mannheim2006alternatives}
Philip~D Mannheim.
\newblock Alternatives to dark matter and dark energy.
\newblock {\em Progress in Particle and Nuclear Physics}, 56(2):340--445, 2006.

\bibitem{bai2024correlators}
Yao Bai, Ting-Long Feng, Suro Kim, Cheng-Yang Lee, Lei-Hua Liu, Wangping Zhao,
  and Siyi Zhou.
\newblock Correlators for pseudo hermitian systems.
\newblock {\em Journal of High Energy Physics}, 2024(11):161, 2024.

\bibitem{sarathi2021application}
Partha Sarathi and Nilesh~Kumar Pathak.
\newblock Application of exact solution of complex morse potential to
  investigate physical systems with complex and negative masses.
\newblock {\em Journal of Physics Communications}, 5(6):065006, 2021.

\bibitem{sarathi2025exact}
Partha Sarathi and Bhaskar~Singh Rawat.
\newblock Exact solution of schr{\"o}dinger equation for the complex morse
  potential to investigate physical systems with position-dependent complex
  mass.
\newblock {\em Physica Scripta}, 100(8):085259, 2025.

\bibitem{andrianov2006complex}
AA~Andrianov, F~Cannata, and A~Yu Kamenshchik.
\newblock Complex lagrangians and phantom cosmology.
\newblock {\em Journal of Physics A: Mathematical and General},
  39(32):9975--9982, 2006.

\bibitem{de2009non}
Paulo Eduardo~Gon{\c{c}}alves de~Assis.
\newblock {\em Non-Hermitian Hamiltonians in field theory}.
\newblock PhD thesis, City University, 2009.

\bibitem{gomes2024non}
YMP Gomes.
\newblock Non-hermitian dirac theory from lindbladian dynamics.
\newblock {\em The European Physical Journal C}, 84(9):958, 2024.

\bibitem{ohlsson2016non}
Tommy Ohlsson.
\newblock Non-hermitian neutrino oscillations in matter with pt symmetric
  hamiltonians.
\newblock {\em Europhysics Letters}, 113(6):61001, 2016.

\bibitem{chernodub2025pseudoreal}
Maxim~N Chernodub, Peter Millington, and Esra Sablevice.
\newblock Pseudoreal quantum fields.
\newblock {\em Physical Review D}, 112(6):065007, 2025.

\bibitem{rizzo2025toward}
Thomas~G Rizzo.
\newblock Toward uv models of kinetic mixing and portal matter. vii. a light
  dark photon in the 3 c 3 l 1 a 1 b model.
\newblock {\em Physical Review D}, 111(7):075018, 2025.

\bibitem{mcdonald2020exponentially}
Alexander McDonald and Aashish~A Clerk.
\newblock Exponentially-enhanced quantum sensing with non-hermitian lattice
  dynamics.
\newblock {\em Nature communications}, 11(1):5382, 2020.

\end{thebibliography}
		
	\end{document}